\documentclass[aps,prb,twocolumn,groupedaddress,showpacs]{revtex4} 
\usepackage{graphicx} 
\usepackage{subfig}
\usepackage{natbib}

\begin{document}  
  
\title{Computational searches for iron oxides at high pressures}

\author{Gihan L.\ Weerasinghe}
  
\affiliation {Theory of Condensed Matter Group, Cavendish Laboratory,
  J J Thomson Avenue, Cambridge CB3 0HE, United Kingdom}
  
\author{Chris J.\ Pickard}

\affiliation {Department of Physics and Astronomy, University College
  London, Gower Street, London WC1E 6BT, United Kingdom}
 
\author{R.\ J.\ Needs}
  
\affiliation {Theory of Condensed Matter Group, Cavendish Laboratory,
  J J Thomson Avenue, Cambridge CB3 0HE, United Kingdom}

\date{\today}   

\begin{abstract}  
  We have used density-functional-theory methods and the \textit{ab
    initio} random structure searching (AIRSS) approach to predict
  stable structures and stoichiometries of mixtures of iron and oxygen
  at high pressures.  Searching was performed for 12 different
  stoichiometries at pressures of 100, 350 and 500 GPa, which involved
  relaxing more than 32,000 structures.  We find that Fe$_2$O$_3$ and
  FeO$_2$ are the only phases stable to decomposition at 100 GPa,
  while at 350 and 500 GPa several stoichiometries are found to be
  stable or very nearly stable.  We report a new structure of
  Fe$_2$O$_3$ with $P2_12_12_1$ symmetry which is found to be more
  stable than the known Rh$_2$O$_3$(II) phase at pressures above
  $\sim$233 GPa.  We also report two new structures of FeO, with
  $Pnma$ and $R\bar{3}m$ symmetries, which are found to be stable
  within the ranges 195--285 GPa and 285--500 GPa, respectively, and
  two new structures of Fe$_3$O$_4$ with $Pca2_1$ and $P2_1/c$
  symmetries, which are found to be stable within the ranges 100--340
  GPa and 340--500 GPa, respectively.  Finally, we report two new
  structures of Fe$_4$O$_5$ with $P4_2/n$ and $P\bar{3}m1$ symmetries,
  which are found to be stable within the ranges 100--231 GPa and
  231--500 GPa, respectively.  Our new structures of Fe$_3$O$_4$ and
  Fe$_4$O$_5$ are found to have lower enthalpies than their known
  structures within their respective stable pressure ranges.
\end{abstract}
  

%
%
%
%
%
%
%

\maketitle


\section{Introduction}  

Fe/O is an important planetary mineral and a significant component of
the Earth's mantle, and it may also be important within the Earth's
core.  Seismological studies have shown that the density of the
Earth's solid inner core is 2--3\% too low to be plausibly explained
by pure Fe, leading to the suggestion that light elements must also be
present.\cite{Stixrude_2001_elasticity_iron,
  BJWood_1993,Stixrude_1997_inner_core_composition,Alfe_2002,Gillan_2006}
The identities and concentrations of light elements present in the
core have yet to be established, although various arguments favour O,
Si, C, S and H as the most likely candidates.\cite{Fe3CSoundVel}

In an earlier paper we investigated the formation of iron carbides at
high pressures,\cite{FeCAIRSS} and here we study iron oxides.  O has
been suggested as an important light element for explaining both the
density jump at the Earth's inner core boundary and the observed
seismic wave velocities.\cite{ozawa_feo} The possible presence of O
within the core has stimulated a range of experimental
\cite{ozawa_feo,Ozawa_B1-B8,Fischer_FeO,Murakami_FeO} and theoretical
\cite{oxygen_alfe_price,Alfe_2000_Fe/O_solid_solution} investigations
of the Fe/O system, albeit only for a handful of stoichiometries.  In
this paper we present a broader investigation spanning 12 different
stoichiometries at three different pressures.  We search for stable
structures at each stoichiometry and pressure, and construct convex
hull diagrams to determine which phases are stable against
decomposition.  We also present results for transitions between solid
phases up to pressures of 500 GPa.

Temperatures within the Earth extend to above 6000 K
\cite{Anzellini_earth_core}.  Atoms such as O could be present in the
Earth's core as isolated impurities, as in a solid solution or,
alternatively, Fe/O compounds could be formed.  A solid solution has a
much higher configurational entropy than an ordered Fe/O compound plus
crystalline Fe, but the formation of crystalline Fe/O compounds could
lead to more favorable bonding arrangements than are obtained in a
solid solution.  To help understand the conditions under which solid
solutions and ordered compounds might be stable it is necessary to
determine which compounds are energetically favorable, taking into
account a wide range of stoichiometries.  This is a formidable task in
its own right, although it should be augmented by calculating finite
temperature effects arising from the vibrational motion of the atoms,
the configurational entropy of the solid solution, and the electronic
entropy.  We leave the estimation of finite temperature effects to
subsequent work.

We have performed searches at 100, 350 and 500 GPa.  The pressure at
the bottom of the Earth's mantle is about 136 GPa, and our data at 100
GPa correspond to a pressure within the lower mantle.  The pressures
within the solid inner core are believed to be in the range 330--360
GPa,\cite{BJWood_1993,Stixrude_1997_inner_core_composition,Alfe_2002,Gillan_2006}
and our results at 350 GPa are therefore representative of inner core
pressures.  The highest pressure at which we have performed searches
of 500 GPa is above that found within the Earth, although such
pressures occur in other planets, including exoplanets.

\section{Computational Details}

Determining the most stable structure of a material corresponds to
finding the global minimum energy state within a high-dimensional
space. A number of methods have been developed to tackle this problem,
which have met with varying degrees of success.  The method we use,
which has been shown to be successful in determining structures that
have subsequently been verified by experiments
\cite{RSS1,RSS2,Fortes_ammonia_monohydrate_II_2009,Griffiths_Ammonia_Dihydrate_II_2012},
is called \textit{ab initio} random structure searching
(AIRSS).\cite{Pickard_2011_AIRSS} AIRSS has been applied to many
systems at high pressures, including iron \cite{TPa_iron_2009} and
oxygen.\cite{TPa_oxygen_2012} Within AIRSS the energy landscape is
sampled by generating random structures.  The structures are then
relaxed to the local enthalpy minimum for some chosen pressure using
the \textsc{castep} plane-wave density functional theory (DFT)
code.\cite{ClarkSPHPRP05} This process is repeated until the lowest
enthalpy structure is found several times or the available
computational resources have been exhausted.

To improve the efficiency of the searches we can impose a number of
biases.  One bias we use is to reject initial structures in which the
smallest inter-atomic separation is less than some chosen value.
Structures which contain atoms that are very close together are
extremely high in enthalpy and may even cause the geometry
optimization procedure to fail.  Since the number of local minima is
known to increase exponentially with the system size \cite{StillExp},
an additional bias may be required for searches over larger unit
cells.  We use a bias that exploits the fact that low enthalpy
structures tend to possess symmetry.  We can create an initial
structure with a particular symmetry by first selecting a ``structural
unit'' containing $A$ formula units (fu).  This structural unit is
chosen to be the lowest enthalpy structure found in an $A$ fu search.
A random space group containing $B$ symmetry operations is then chosen
and applied to the structural unit. This generates a larger initial
structure containing $A \times B$ fu.  In this investigation $A$ and
$B$ range from 1--3 and 2--4, respectively.

Structures found to be metastable at one pressure may become stable at
another. Expanding the enthalpy around the pressure $p_0$ at which a
search is performed gives
\begin{equation}
\label{taylorenth}
H(p) = H(p_0)+(p-p_0)\frac{dH}{d p} \bigg |_{p_0} + 
\frac{1}{2}(p-p_0)^2  \frac{d^2 H}{d p^2} \bigg |_{p_0} ...
\end{equation}
The second-order derivative in Eq.\ (\ref{taylorenth}) is related to
the bulk modulus and is computationally expensive to evaluate.  A
useful approximation can, however, often be obtained by neglecting the
second-order term and using the fact that the first derivative of $H$
with respect to pressure $p$ at $p_0$ is equal to the volume $v_0$, so
that
\begin{equation}
\label{linearapprox}
H(p) \approx H(p_0)+v_0(p-p_0).
\end{equation}
Since $H(p_0)$ and $v_0$ are evaluated within the structure searching,
Eq.\ (\ref{linearapprox}) may be used to estimate the enthalpies of
the phases at any pressure $p$.  This allows us to determine
approximate pressures at which phase transitions may occur, which we
then refine by performing higher accuracy DFT calculations.

We used ultrasoft pseudopotentials \cite{Vanderbilt90} to represent
the cores of the Fe and O atoms.  For the structure searching we used
a fairly soft Fe pseudopotential in which only the $4s$ and $3d$
electrons were treated explicitly, while for O we treated the $2s$ and
$2p$ electrons explicitly.  For the higher quality calculations we
used a harder Fe pseudopotential in which the $3s$, $3p$, $4s$, and
$3d$ electrons were treated explicitly.  Both the harder Fe
pseudopotential and the O pseudopotential have been successfully
tested up to terapascal
pressures.\cite{TPa_iron_2009,TPa_CO_2011,TPa_oxygen_2012}

The final structures were obtained using a two-step procedure.  We
first used AIRSS in medium-quality calculations which were optimized
for computational speed.  The medium quality settings consisted of a
$k$-point sampling grid of spacing of $2 \pi \times$0.07 \AA$^{-1}$
and a plane-wave cut-off energy of 490 eV.  The lowest enthalpy
structures for each stoichiometry were then further relaxed in a
higher-quality calculation using a finer $k$-point sampling grid of
spacing $2 \pi \times$0.03 \AA$^{-1}$ and a plane-wave cut-off energy
of 1000 eV.  In all of our searches we used the Perdew-Burke-Ernzerhof
(PBE) generalized gradient approximation (GGA) exchange correlation
functional.\cite{PerdewBE96}

The Fe/O system is known to be problematic for DFT-based methods
because of magnetic and strong correlation effects.  The presence of
magnetism in FeO at high pressures has been documented, although there
are some conflicting reports.  M\"{o}ssbauer spectroscopy has shown
that a continuous transition from high-spin to low-spin states
\cite{pasternak_feo} occurs at 90--120 GPa, leading to an eventual
collapse at 140 GPa and 300 K.  This is in contrast to experiments
performed using X-ray emission spectroscopy for FeO at 300 K, which
have suggested that a magnetic state exists up to at least 143 GPa
\cite{badro_feo}, while a magnetic collapse in Fe$_2$O$_3$ has been
observed at 50 GPa.\cite{pasternak_fe2o3} In addition,
zero-temperature static lattice GGA DFT calculations have predicted a
magnetic collapse of FeO at around 200 GPa.\cite{mag_collapse_cohen}
 
The existence of magnetic order under core conditions can be safely
ruled out, as it would be destroyed by the high temperatures.  We have
not performed searches with spin-polarization because they are very
costly, but to investigate the effects of magnetism on the enthalpies,
we have re-relaxed our structures with high quality settings starting
from a high spin state.  If a relaxation did not achieve convergence
we reduced the initial spin state and re-relaxed until convergence was
achieved. We found that the small magnetic moments have a negligible
effect on the enthalpies at the high pressures considered here
($\sim$0.002 eV per atom at 100 GPa in the worst case). For these
reasons we do not believe magnetism plays a significant role at the
pressures we have investigated and hence we have not considered it
further.

Similar arguments can be made about the effects of strong Coulomb
correlations which arise in iron oxides from interactions involving
the localized Fe $d$ electrons.  The strength of such correlations is
conventionally measured by the parameter $U/W$, where $U$ is the
Hubbard $U$ energy and $W$ is the $d$ band width.  Electrons tend to
delocalize at high pressures, which has the effect of increasing $W$
and reducing $U$.\cite{mag_collapse_cohen} Under these conditions,
where the correlations are less strong, GGA DFT is likely to be
reasonably accurate.\cite{FeO_FP_HP}

\begin{figure}
\centering
 \subfloat[]{
  \label{fig:convexhull:100}
  \includegraphics[width=.8\linewidth]{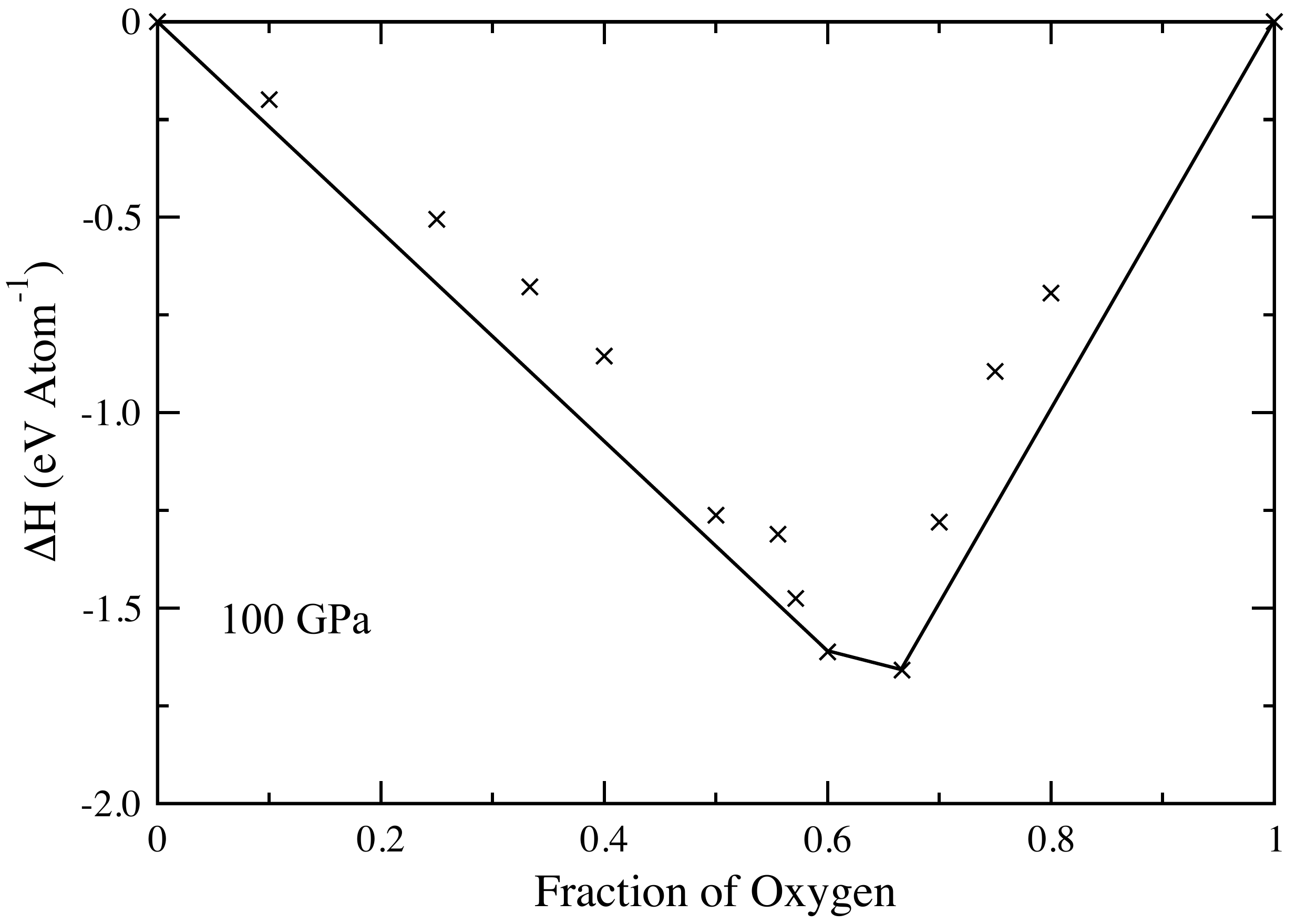} } 
  \\
  \subfloat[]{
  \label{fig:convexhull:350}
  \includegraphics[width=.8\linewidth]{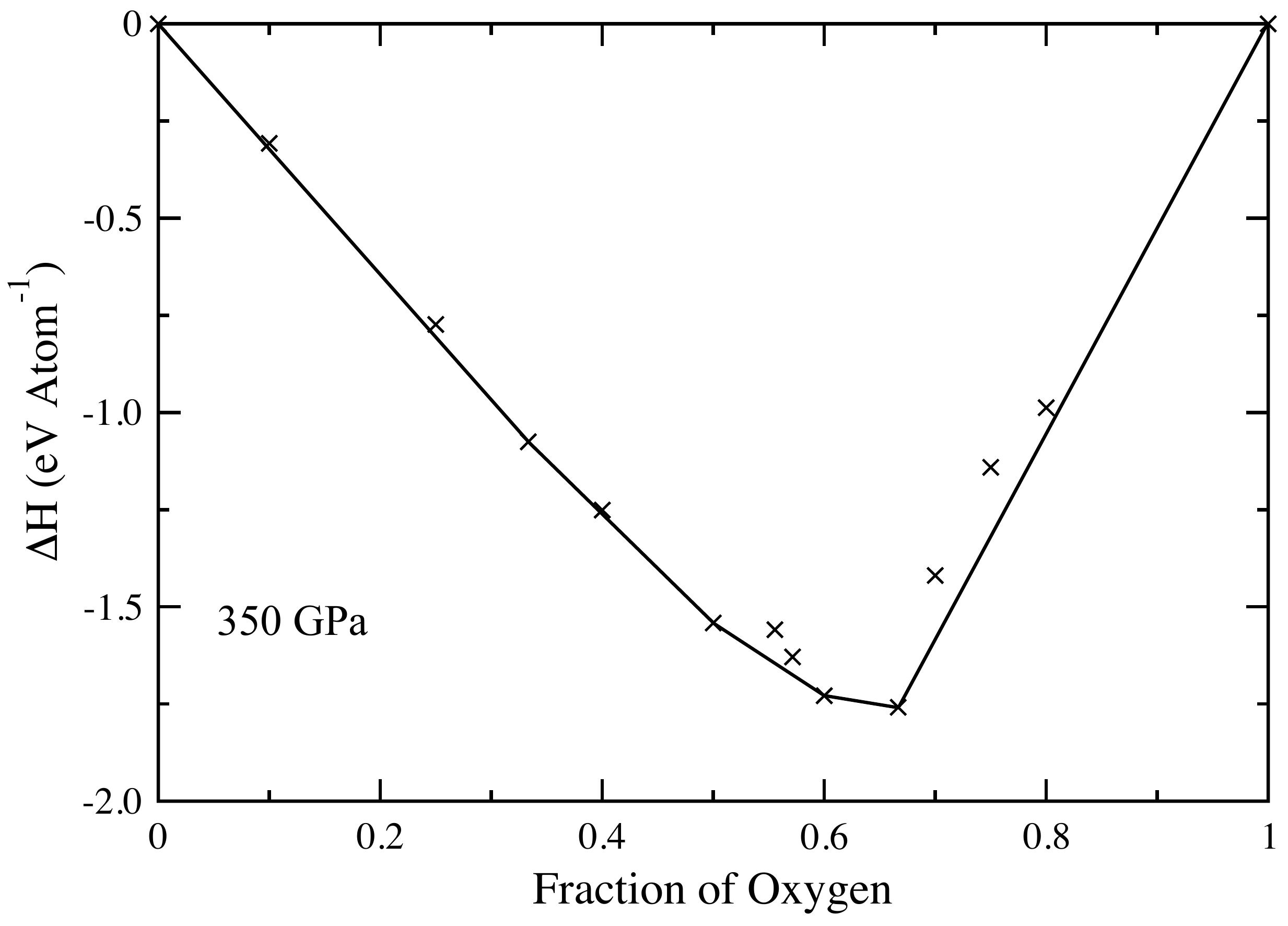} }
  \\
  \subfloat[]{
  \label{fig:convexhull:500}
  \includegraphics[width=.8\linewidth]{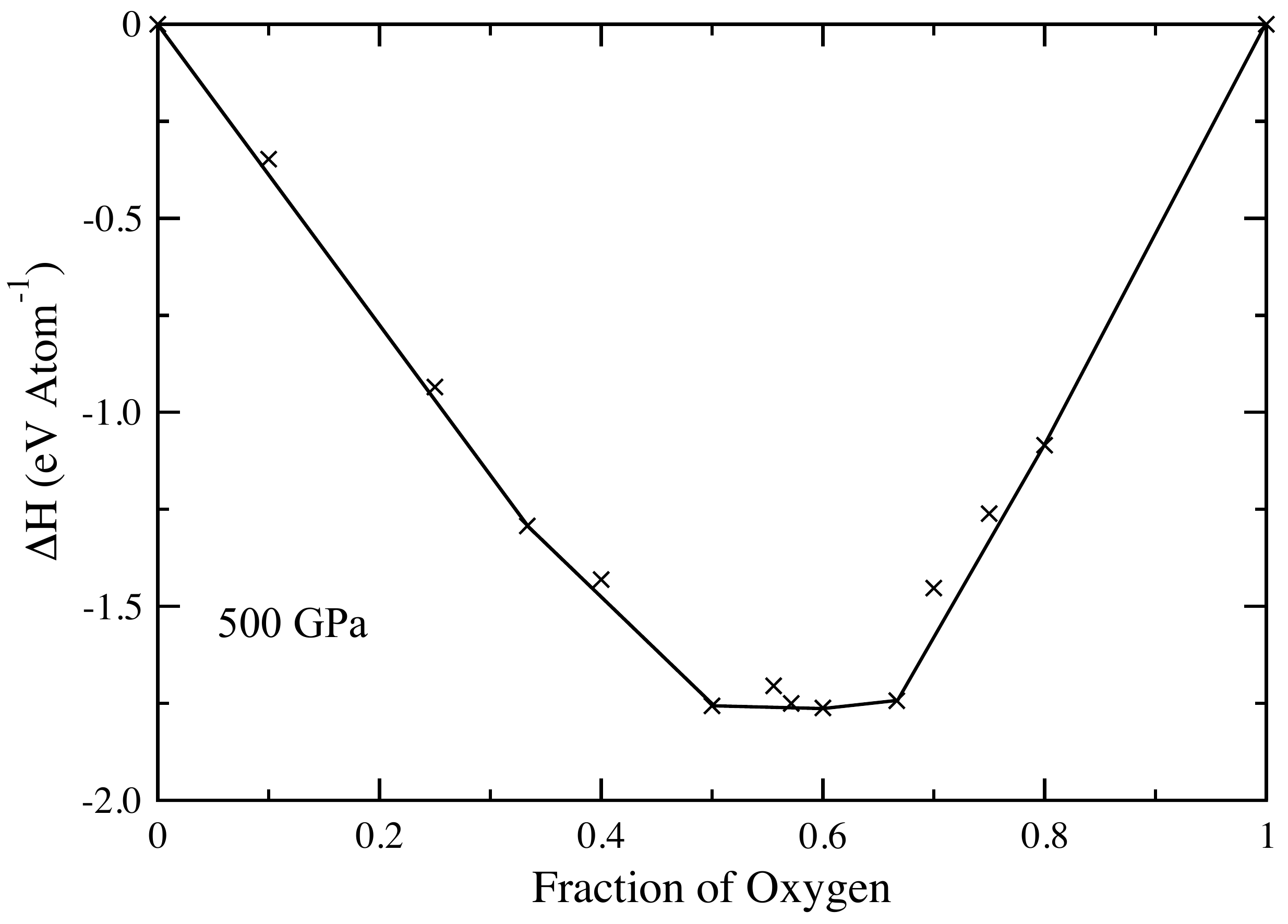} }
\caption{\label{fig:convexhull} Convex hull diagrams
  for Fe/O at 100, 350 and 500 GPa.}
\end{figure}

\begin{table} 
\begin{ruledtabular} 
\begin{tabular}{lcccccccc} &$X$&$N$ &$S_{100}$ &  $S_{350}$&  $S_{500}$  \\
\hline 
Fe$_9$O &0.10&3&[$Amm2 \ (1)]$&[$Amm2 \ (1)]$&[$Amm2 \ (1)]$  \\ 
Fe$_3$O &0.25&4&[$P\bar{6}m2 \ (1)]$&[$P\bar{6}m2 \ (1)] $&[$P4/nmm \ (2)$]  \\
Fe$_2$O &0.33&6&$P6_3/mmc \ (2)$&$I4/mmm \ (1)$&$I4/mmm \ (1)$  \\              
Fe$_3$O$_2$ &0.40&4&$Pm \ (3)$&$C2/m \ (1)$&[$Pm \ (2)$] \\
FeO &0.50&8&$P3_221 \ (3)$&$R\bar{3}m \ (1)$&$R\bar{3}m \ (1)$ \\
Fe$_4$O$_5$ &0.56&2&$P4_2/n \ (2)$&$P\bar{3}m1 \ (1)$&$P\bar{3}m1 \ (1)$ \\
Fe$_3$O$_4$ &0.57&4&$Pca2_1 \ (4)$&$P2_1/c \ (4)$&$P2_1/c \ (4)$ \\
Fe$_2$O$_3$ &0.60&4&$Pbcn \ (4)$&$P2_12_12_1 \ (4)$&$P2_12_12_1 \ (4)$\\
FeO$_2$&0.66&5&$Pa\bar{3} \ (4)$&$Pa\bar{3} \ (4)$&$R\bar{3}m \ (1)$ \\
Fe$_3$O$_7$&0.70&3&$P2/m \ (1)$&$I\bar{4}3d \ (2)$&$I\bar{4}3d \ (2)$ \\
FeO$_3$&0.75&4&$P\bar{1} \ (4)$&$P\bar{1} \ (4)$&$Cmcm \ (2)$ \\
FeO$_4$&0.80&4&$Fdd2 \ (2)$&$P2_1/c \ (2)$&$P2_1/c \ (2)$ 
\end{tabular} 
\end{ruledtabular} 
\caption{\label{table:structures} {
    The lowest enthalpy structures found from searching at 100, 350 and 500 GPa. 
    $X$ is the fraction of oxygen atoms.  Entries which show space groups within square 
    brackets indicate mixtures of structures of phases with other stoichiometries, see text.  
    $N$ is the maximum number of formula units used in the searches.  $S_{P}$ is the space group of 
    the lowest enthalpy structure found at pressure $P$.  In each case, the number of 
    formula units in the primitive cell is given in round brackets.
  }} 
\end{table}

\begin{figure*}
\centering
\subfloat[Fe$_2$O ($I4/mmm$) (350 GPa) ]{
\label{fig:structures_fe2o_i4mmm}
\includegraphics[width=.35\linewidth]{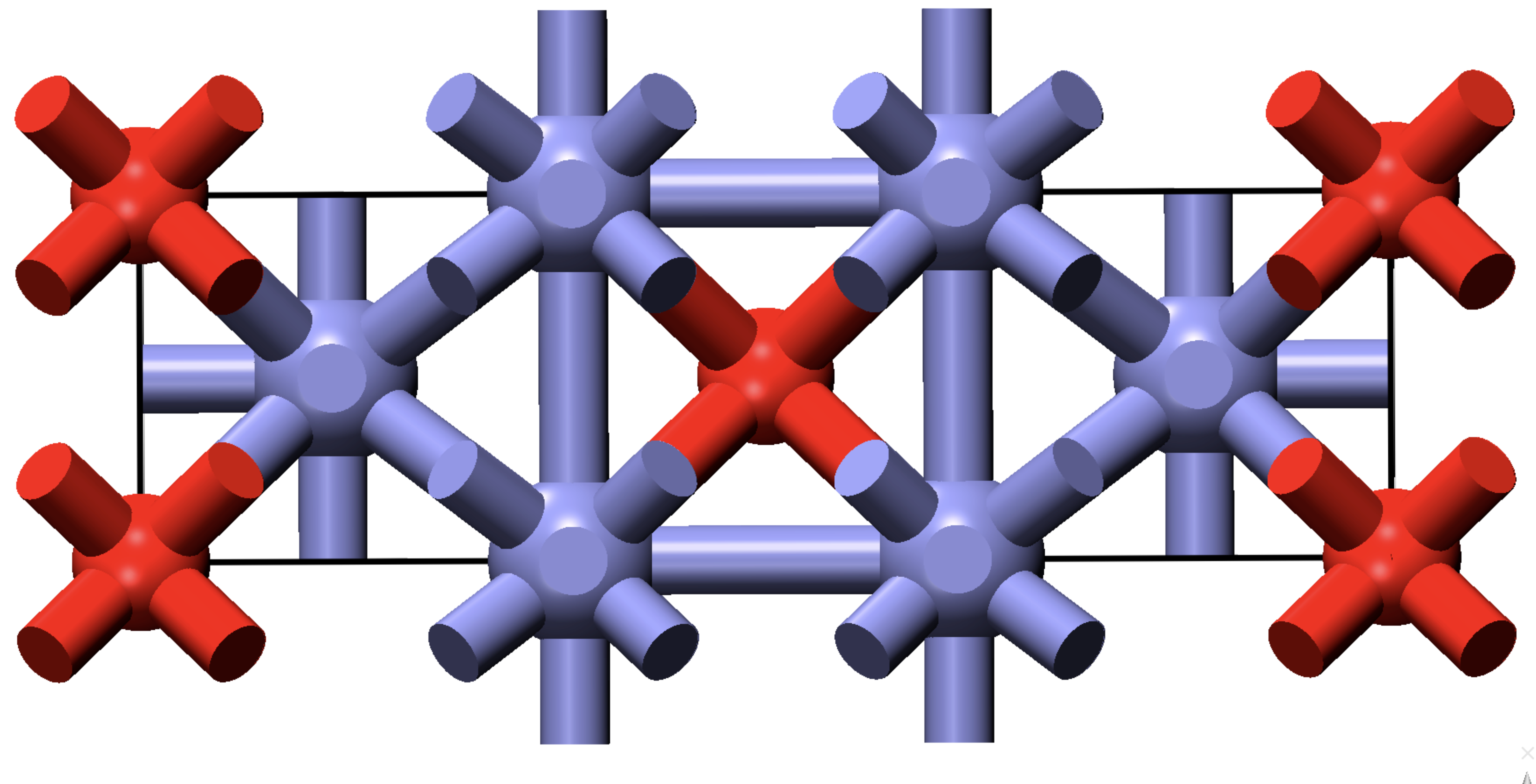} }
\subfloat[FeO ($R\bar{3}m$) (350 GPa)]{
\label{fig:structures_feo_r-3m}
\includegraphics[width=.35\linewidth]{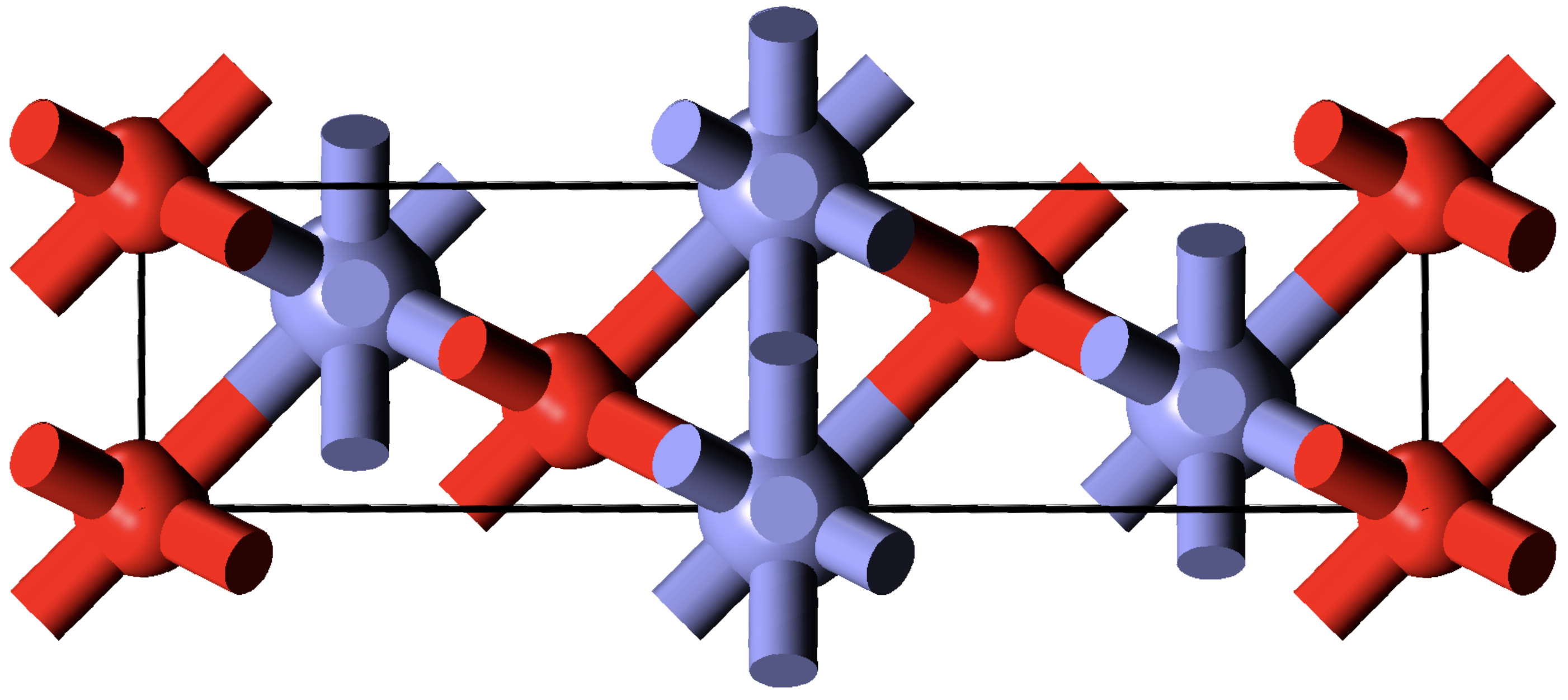} }
\\
\subfloat[Fe$_2$O$_3$ ($P2_12_12_1$) (350 GPa)]{
\label{fig:structures_fe2o3_p212121}
\includegraphics[width=.35\linewidth]{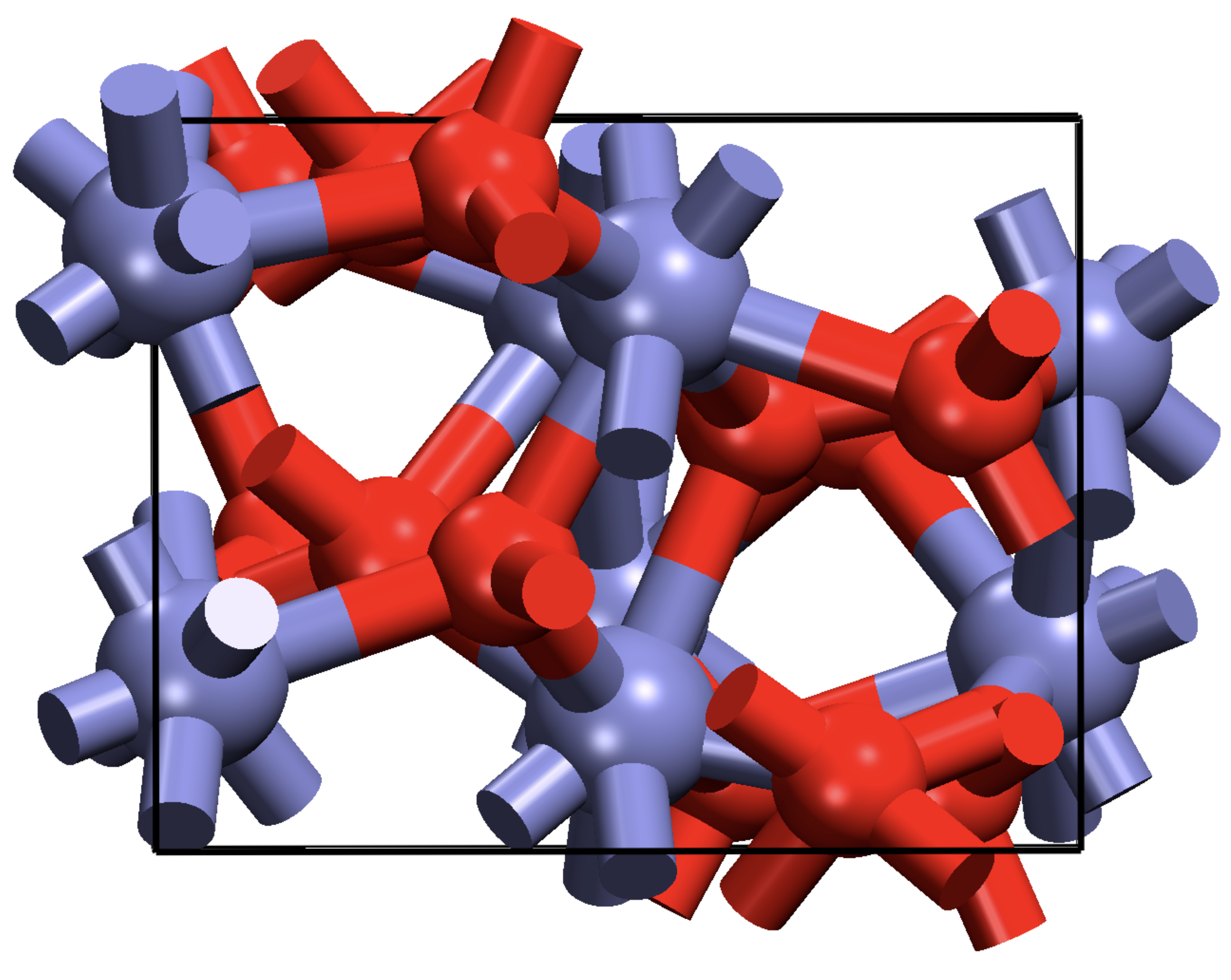} }
\subfloat[FeO$_2$ ($Pa\bar{3}$) (100 GPa)]{
\label{fig:structures_feo2_pa-3}
\includegraphics[width=.31\linewidth]{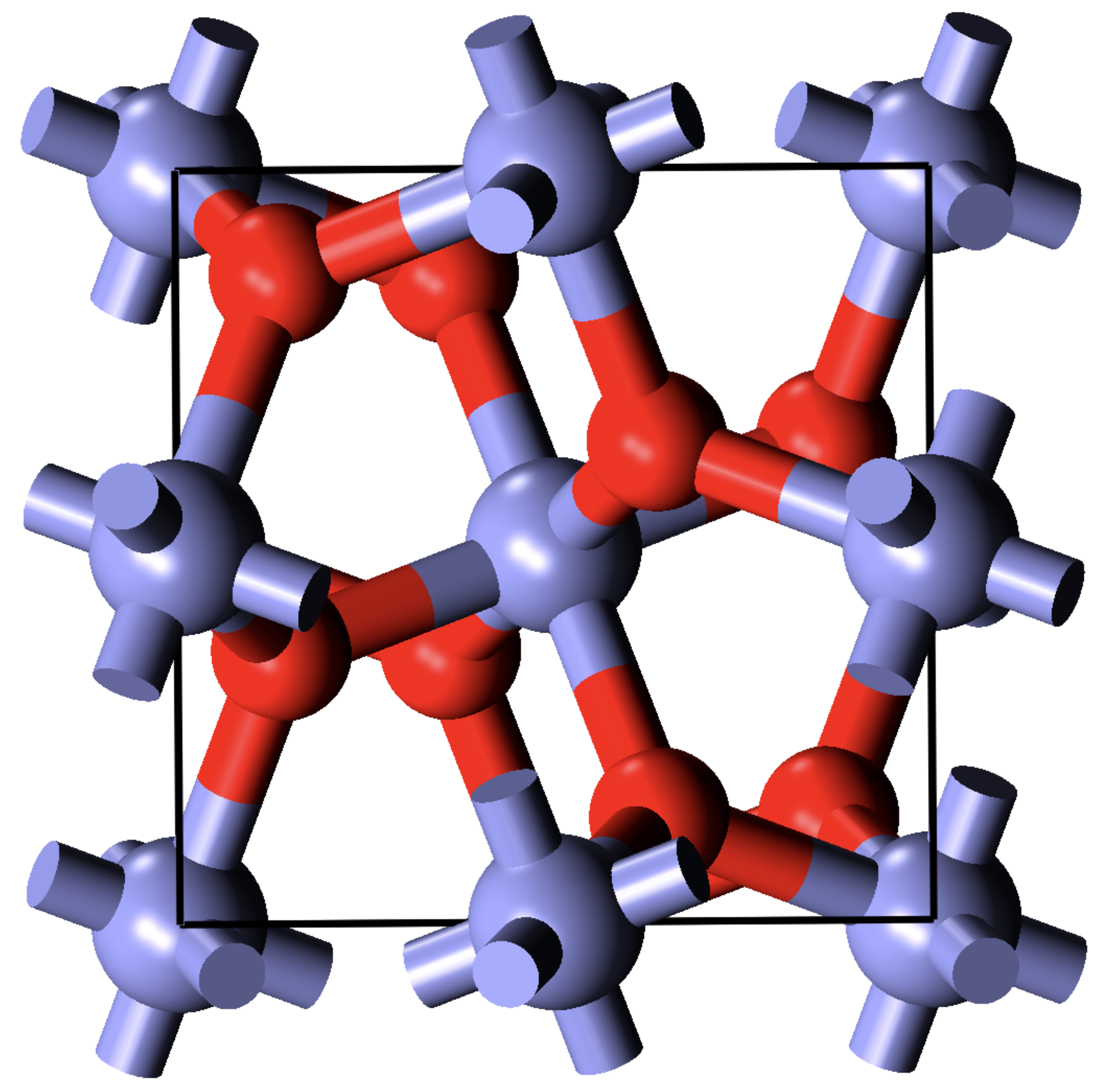} }
\\
\subfloat[FeO$_2$ ($R\bar{3}m$) (500 GPa)]{
\label{fig:structures_feo2_r-3m}
\includegraphics[width=.35\linewidth]{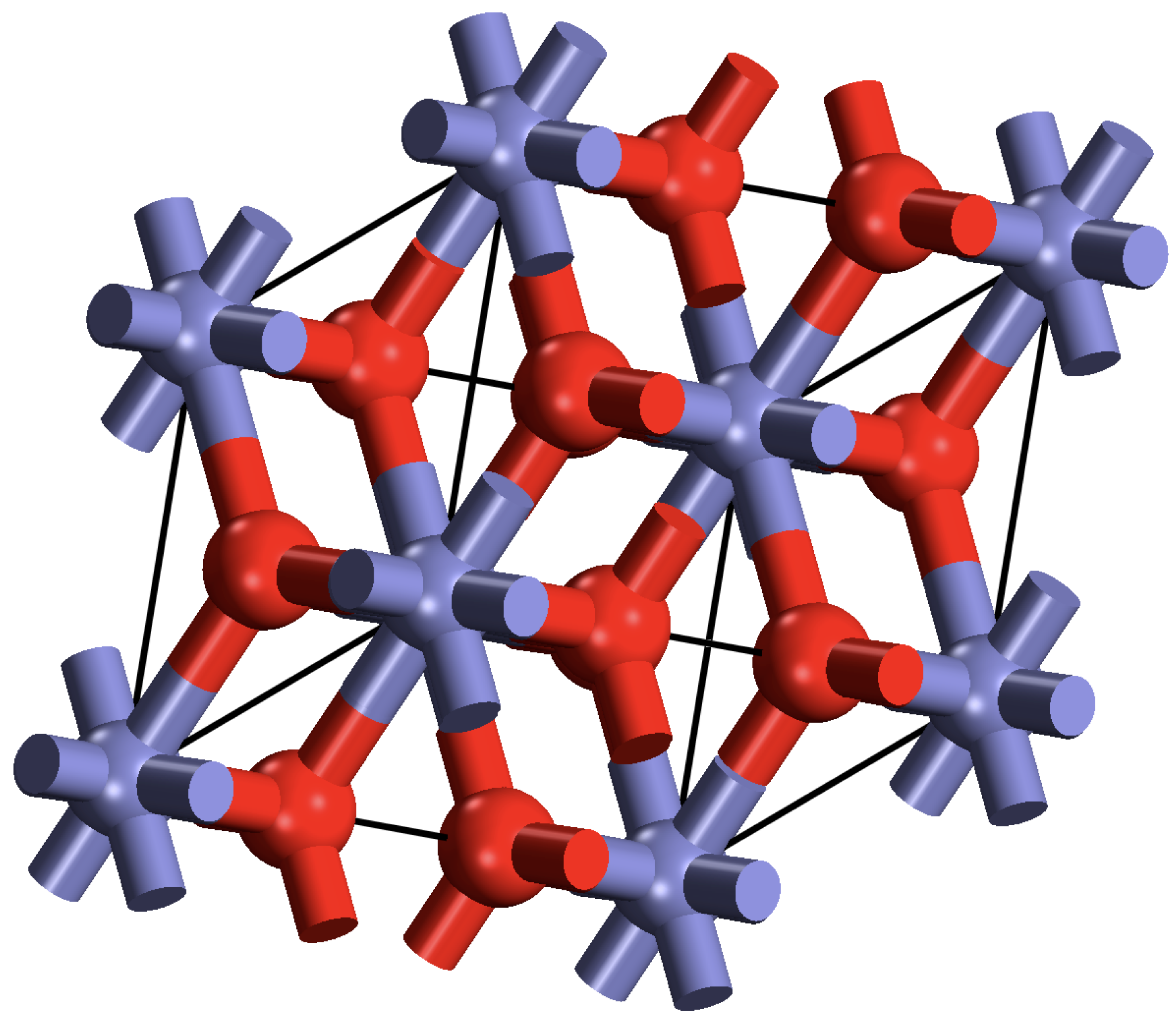} }
\\
\subfloat[FeO$_4$ ($P2_1/c$) (500 GPa)]{
\label{fig:structures_feo4_p21c}
\includegraphics[width=.60\linewidth]{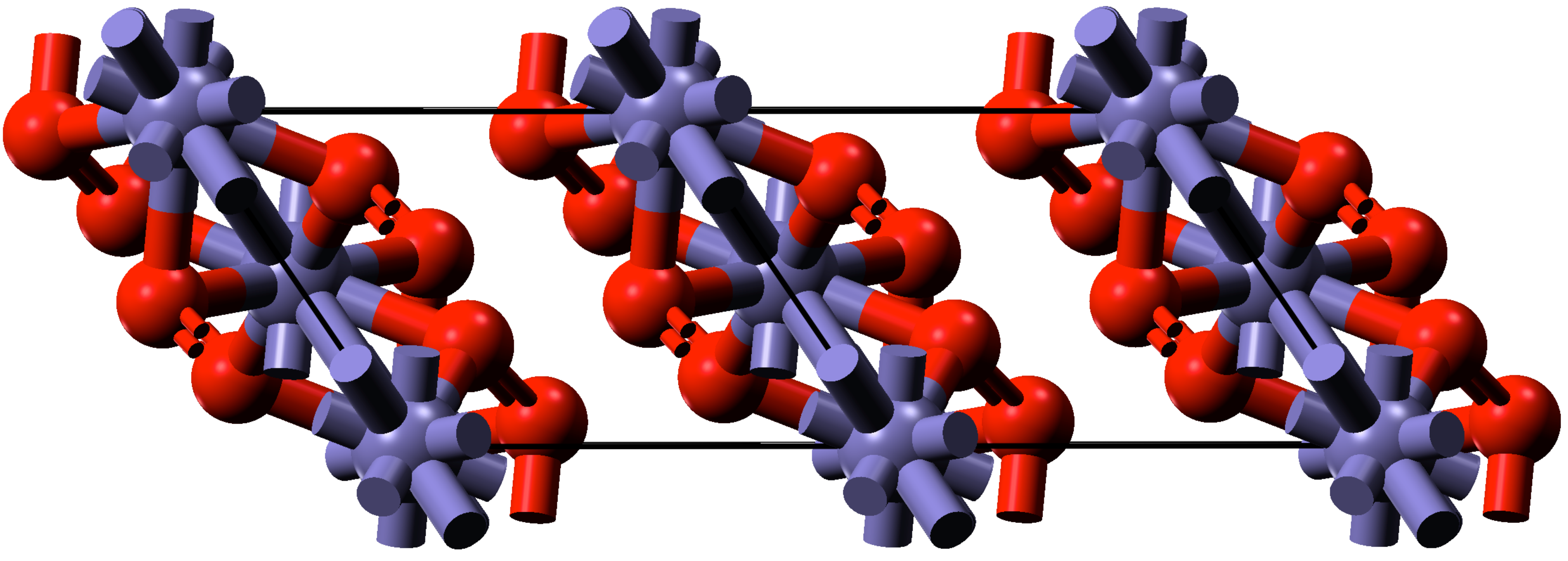} }
\caption{\label{fig:structures}(Color online). Structures found from
  searching that lie on the convex hull. The O atoms are shown in red
  and the Fe atoms are in blue.}
\end{figure*}

\section{Convex Hulls}  

We have plotted our energy data on convex hull diagrams to assess the
stability of the phases with respect to decomposition.  To construct
the hulls we also need the enthalpies of the end members, pure Fe and
pure O.  Fe is predicted to adopt the hexagonal close packed (hcp)
structure over the pressure range of interest here, while for oxygen
at 500 GPa we use the $P6_3/mmc$ phase predicted by Sun \textit{et
  al.}\ \cite{TPa_oxygen_2012} At pressures below 500 GPa we use the
$\zeta$ phase.\cite{zeta_oxygen}

The enthalpy of formation per atom of a compound with respect to its
elements is
\begin{equation}
\label{relenth}
\Delta H_{\rm s}=\frac{H_{\mathrm{s}}-(N_{\mathrm{Fe}}H_{\mathrm{Fe}}+N_{\mathrm{O}}H_{\mathrm{O}})}{N_{\mathrm{O}}+N_{\mathrm{Fe}}},
\end{equation}
where $N_x$ and $H_x$ are the number of atoms and enthalpy per atom of
element $x$, respectively, and $H_{\rm s}$ is the enthalpy of the
structure.  A negative $\Delta H_{\rm s}$ indicates that the structure
is stable with respect to decomposition into its elements, although it
may be unstable with respect to decomposition into compounds of other
stoichiometries.  Structures lying above the convex hull are unstable
with respect to decomposition into other compounds and are therefore
less likely to form.  Structures lying on the convex hull are
thermodynamically stable and are therefore more likely to form.


The convex hull diagrams at 100, 350 and 500 GPa shown in Fig.\
\ref{fig:convexhull} exhibit fairly smooth variations with
composition.  Figure \ref{fig:convexhull}\subref{fig:convexhull:100}
shows that the only stoichiometries that lie on the hull at 100 GPa
are Fe$_2$O$_3$ and FeO$_2$.  At 350 GPa, Fe$_9$O, Fe$_3$O, Fe$_2$O,
Fe$_3$O$_2$, FeO, Fe$_2$O$_3$ and FeO$_2$ are found to lie on, or very
close to the hull, and similar results are found at 500 GPa. Perhaps
the most notable difference at 500 GPa is that FeO$_4$ is found to lie
on the hull.  FeO$_2$ and Fe$_2$O$_3$ are predicted to be stable with
respect to decomposition at all pressures investigated. Fe$_3$O$_4$,
which is believed to be an important phase in the Earth's lower mantle
\cite{dubrovinsky2003structure}, is found to be marginally unstable
with respect to decomposition at the pressures considered here.

Increasing pressure appears to stabilize Fe rich phases, an effect
that is most noticeable between 100 and 350 GPa.  There is also a
stabilization of O rich phases with increasing pressure. For example,
FeO$_3$ and FeO$_4$ can be seen to move considerably closer to the
hull between 100 GPa and 500 GPa.  The former also coincides with a
transition from a low symmetry $P\bar{1}$ phase to a phase of $Cmcm$
symmetry.

It is possible that structures that lie just above the convex hull
correspond to metastable mixtures of phases rather than single phases.
This is particularly likely for large unit cells.  We have found
several such cases in our results, and the fact that they lie just
above the hull suggests that the interfacial energies between the
different structures present are small.  The $\bar{P}6m2$ structure of
Fe$_3$O is found to be a mixture of phases consisting of Fe and
Fe$_2$O while the $P4/nmm$ structure is found to consist of Fe,
Fe$_2$O and FeO phases. Finally, Fe$_9$O ($Amm2$) is found to consist
of Fe and Fe$_2$O phases and the $Pm$ phase of Fe$_3$O$_2$ (at 500
GPa) is found to consist of Fe$_2$O and FeO phases.  These mixtures of
phases are indicated by square brackets in Table
\ref{table:structures}.

\begin{figure*}
\centering
  \subfloat[Fe$_2$O$_3$ ($Pbcn$) at 100 GPa.]{
  \includegraphics[width=.37\linewidth]{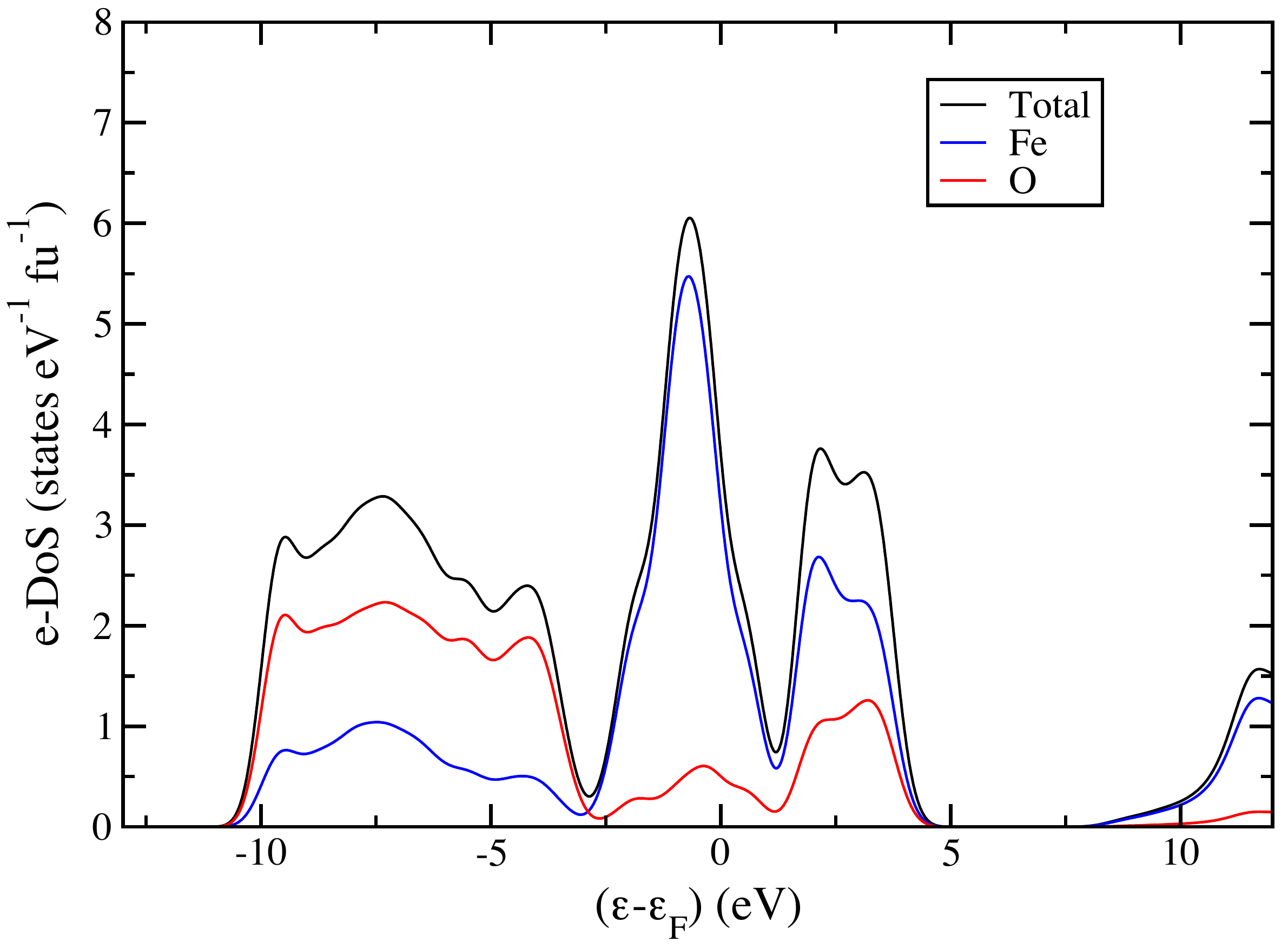} } 
  \subfloat[Fe$_2$O ($I4/mmm$) at 350 GPa.]{
  \includegraphics[width=.37\linewidth]{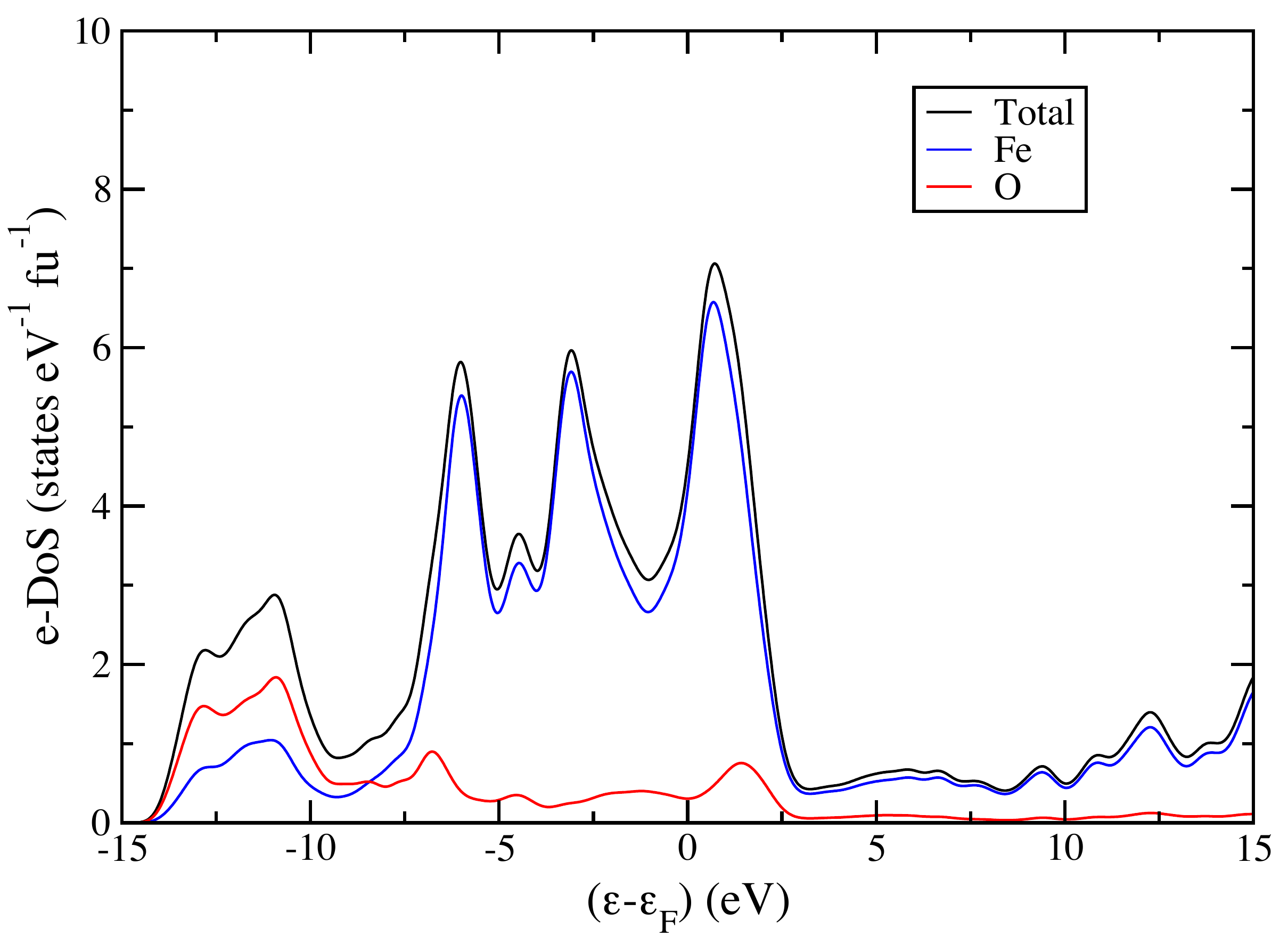} }
  \\
  \subfloat[FeO ($R\bar{3}m$) at 350 GPa.]{
  \includegraphics[width=.37\linewidth]{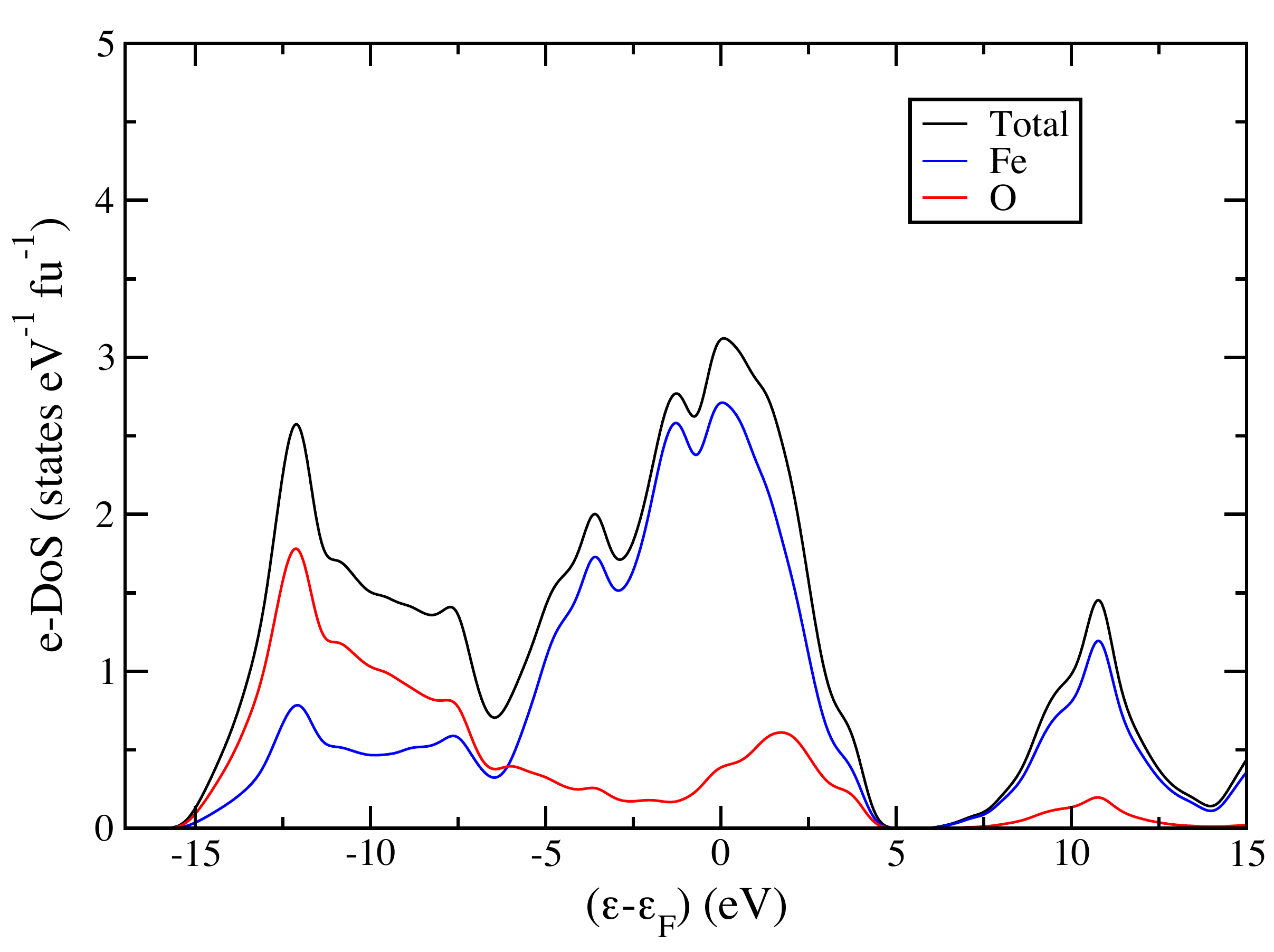} }
  \subfloat[Fe$_2$O$_3$ ($P2_12_12_1$) at 350 GPa.]{
  \includegraphics[width=.37\linewidth]{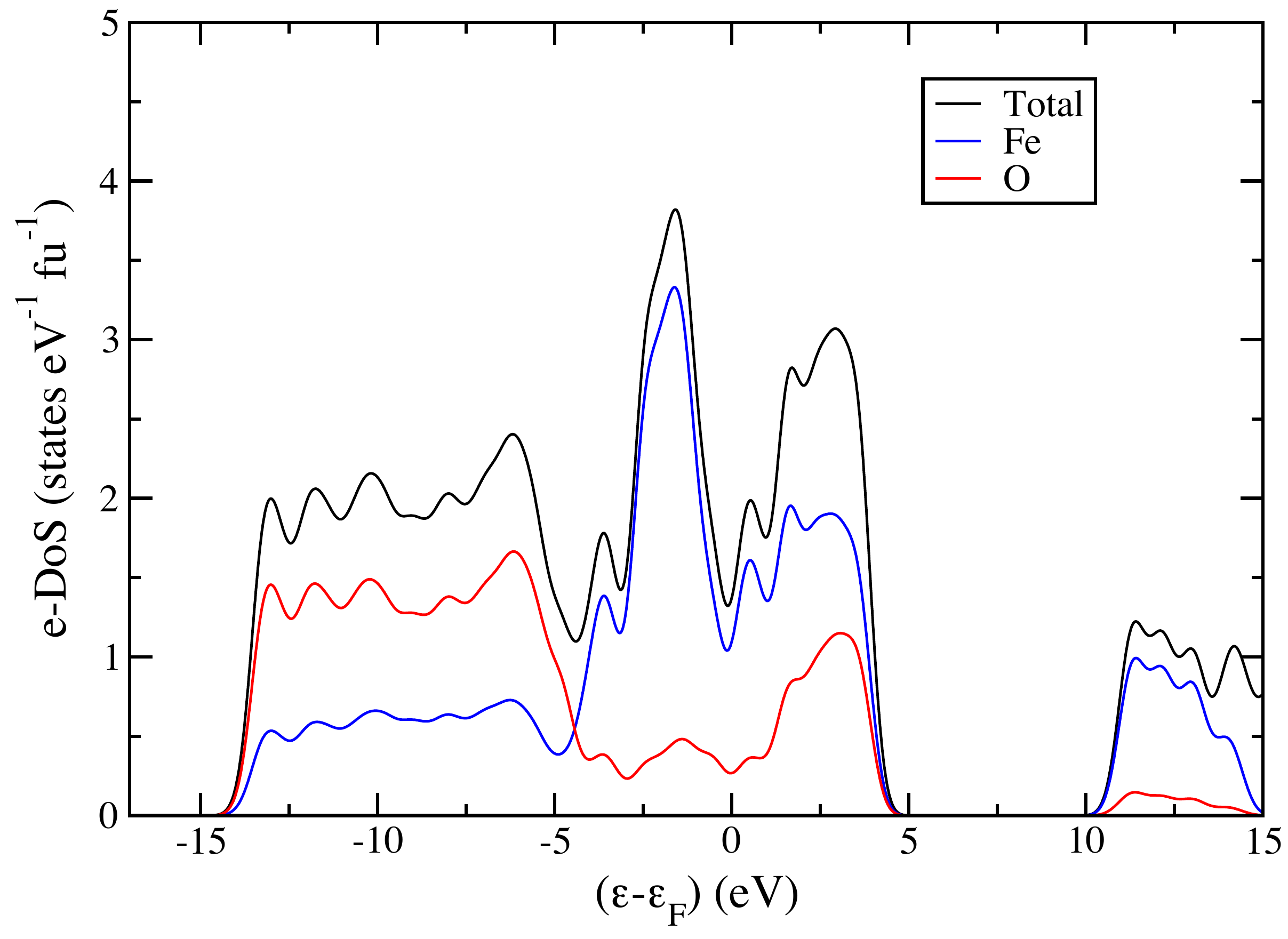} }
  \\
  \subfloat[FeO$_2$ ($Pa\bar{3}$) at 350 GPa.]{
  \includegraphics[width=.37\linewidth]{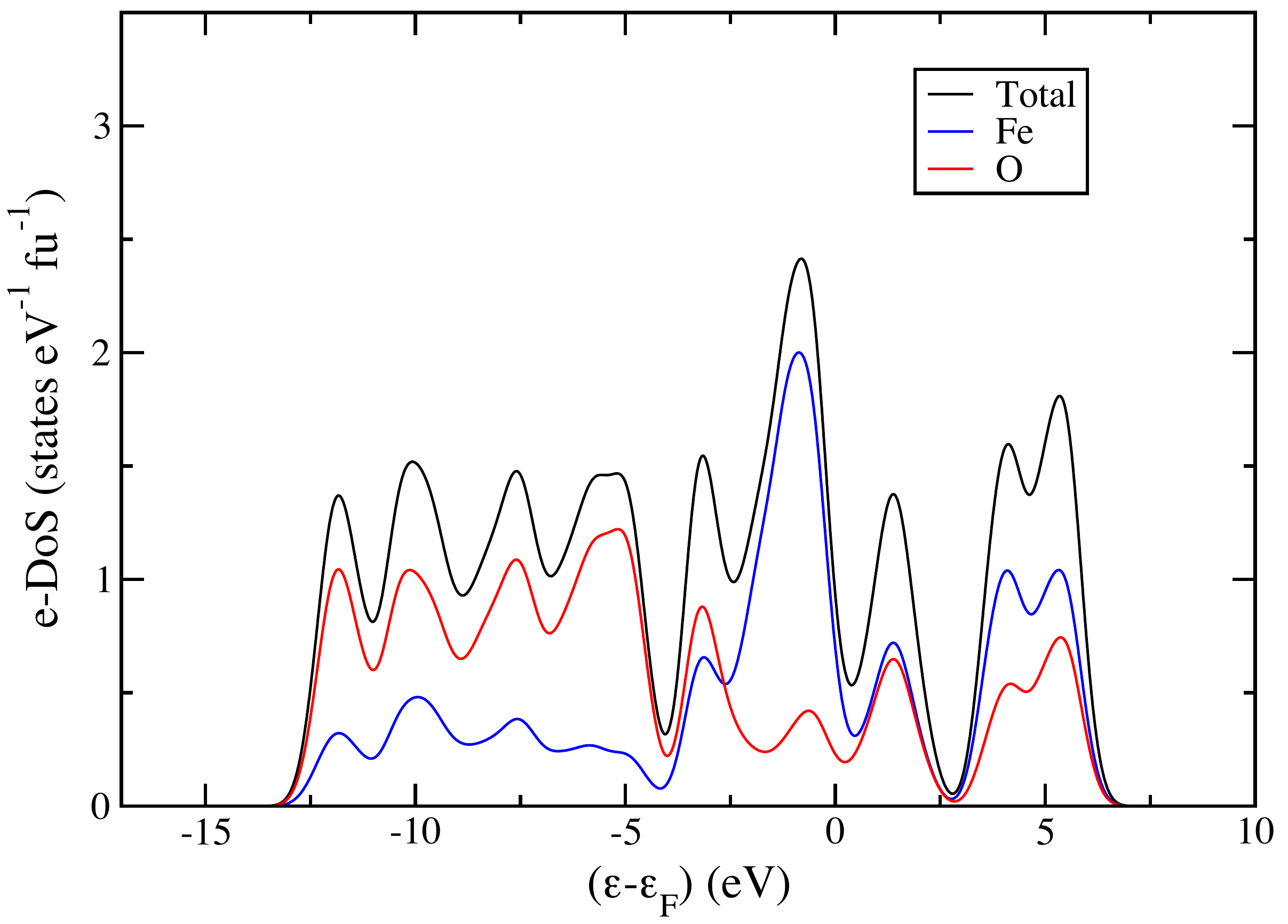} } 
  \subfloat[FeO$_2$ ($R\bar{3}m$) at 500 GPa.]{
  \includegraphics[width=.37\linewidth]{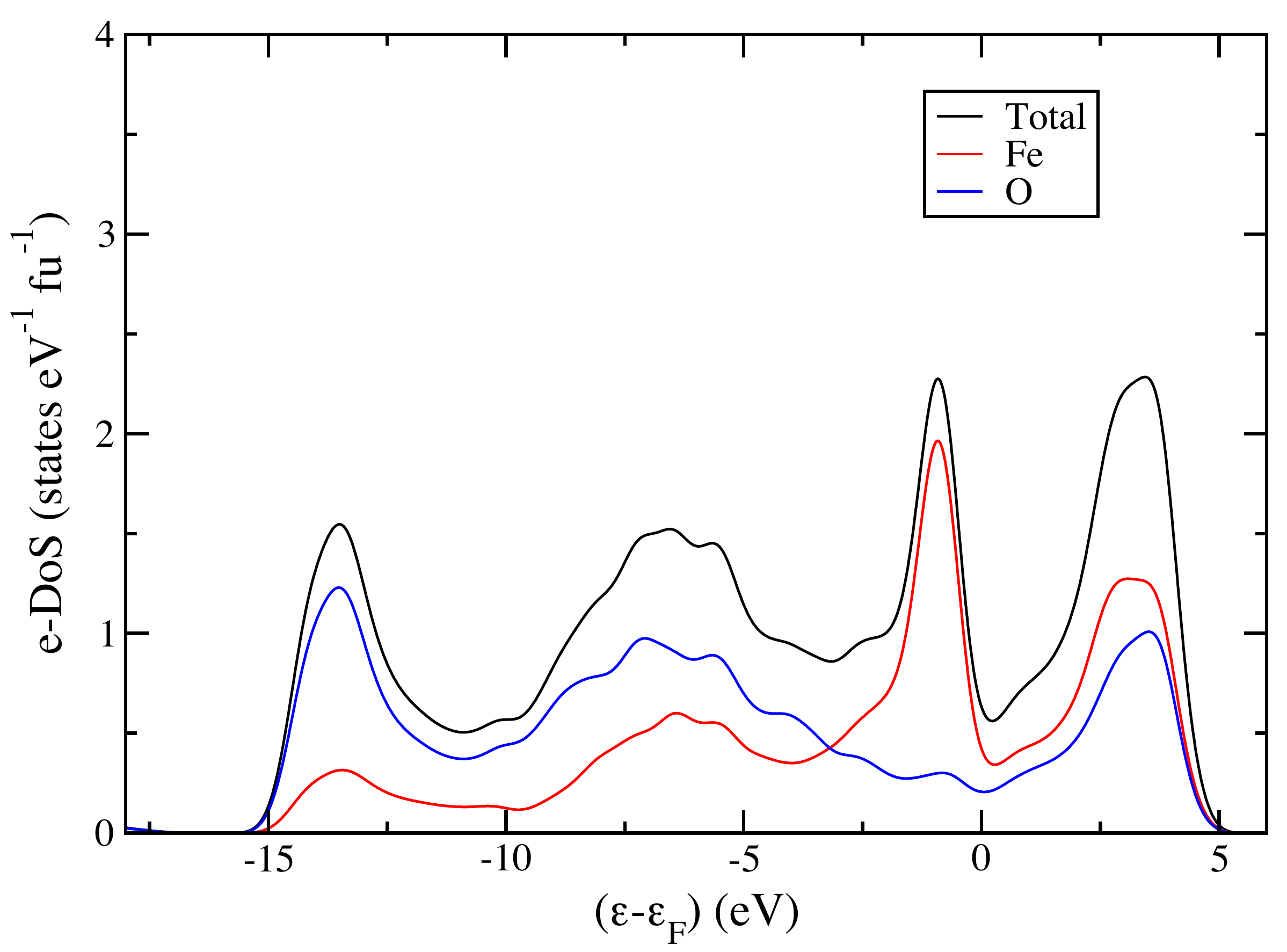} }
  \\
  \subfloat[FeO$_4$ ($P2_1/c$) at 500 GPa.]{
  \includegraphics[width=.37\linewidth]{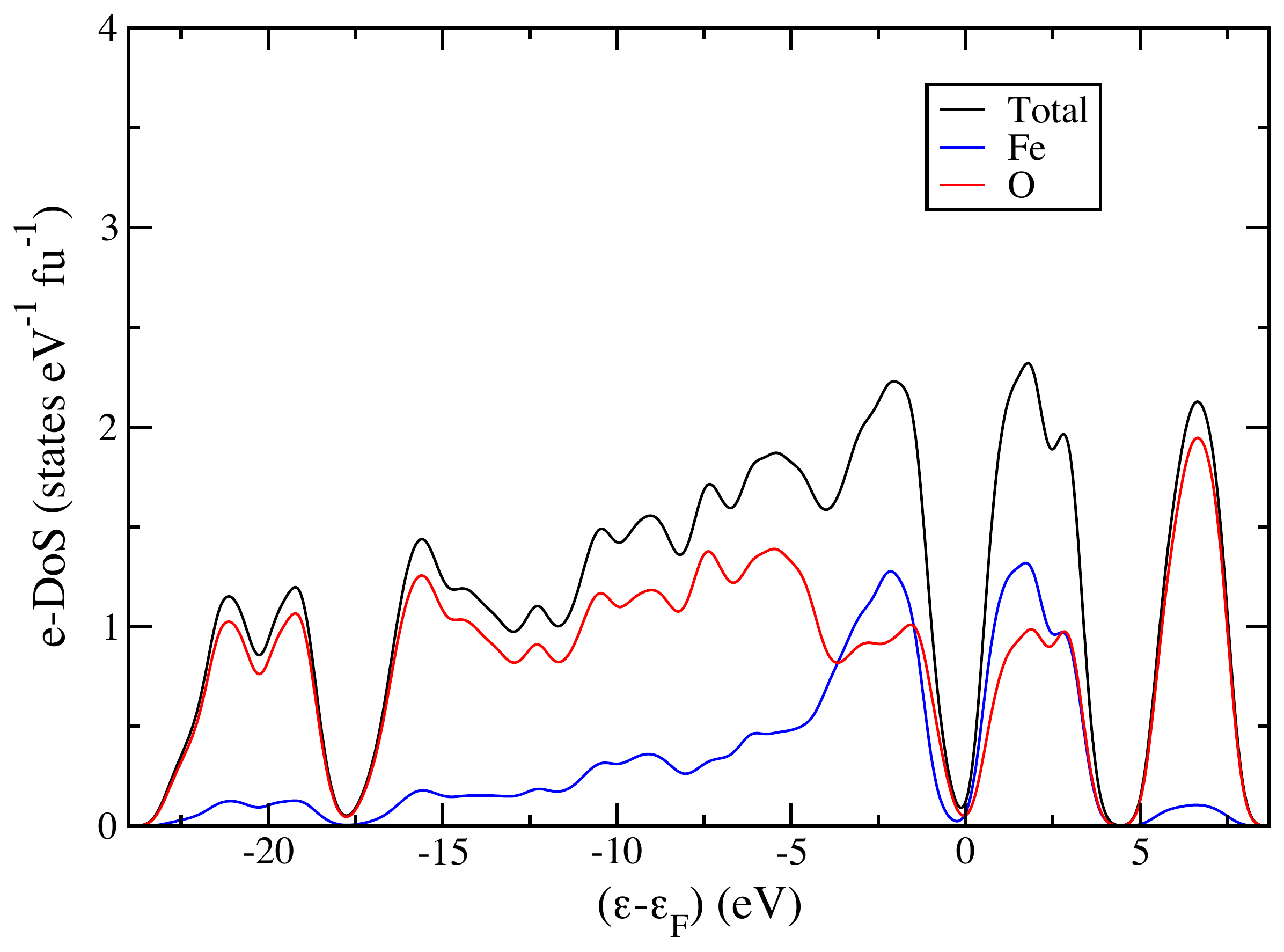} } 
\caption{\label{fig:alldos} (Color online).  Electronic densities of
  states for structures lying on the convex hull.  Each plot shows the
  total density of states and its decomposition into contributions
  from Fe and O.  The Fermi energies are at 0 eV.}
\end{figure*}

\section{Phase transitions} 

\begin{figure*}
\centering

  \subfloat[Fe$_3$O]{
  \label{fig:ptplots:fe3o}
  \includegraphics[width=.31\linewidth]{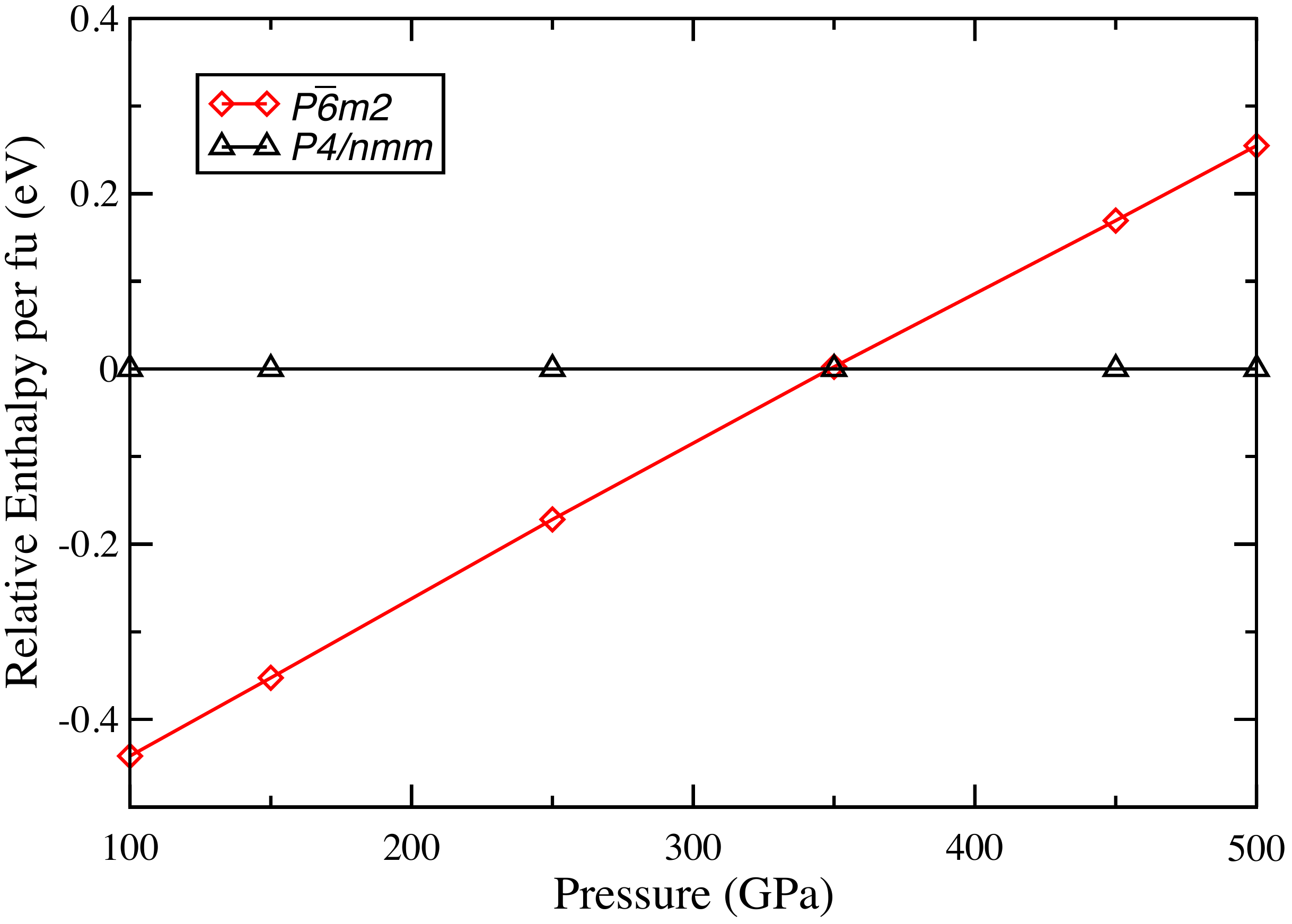} }
  \subfloat[Fe$_2$O]{
  \label{fig:ptplots:fe2o}
  \includegraphics[width=.31\linewidth]{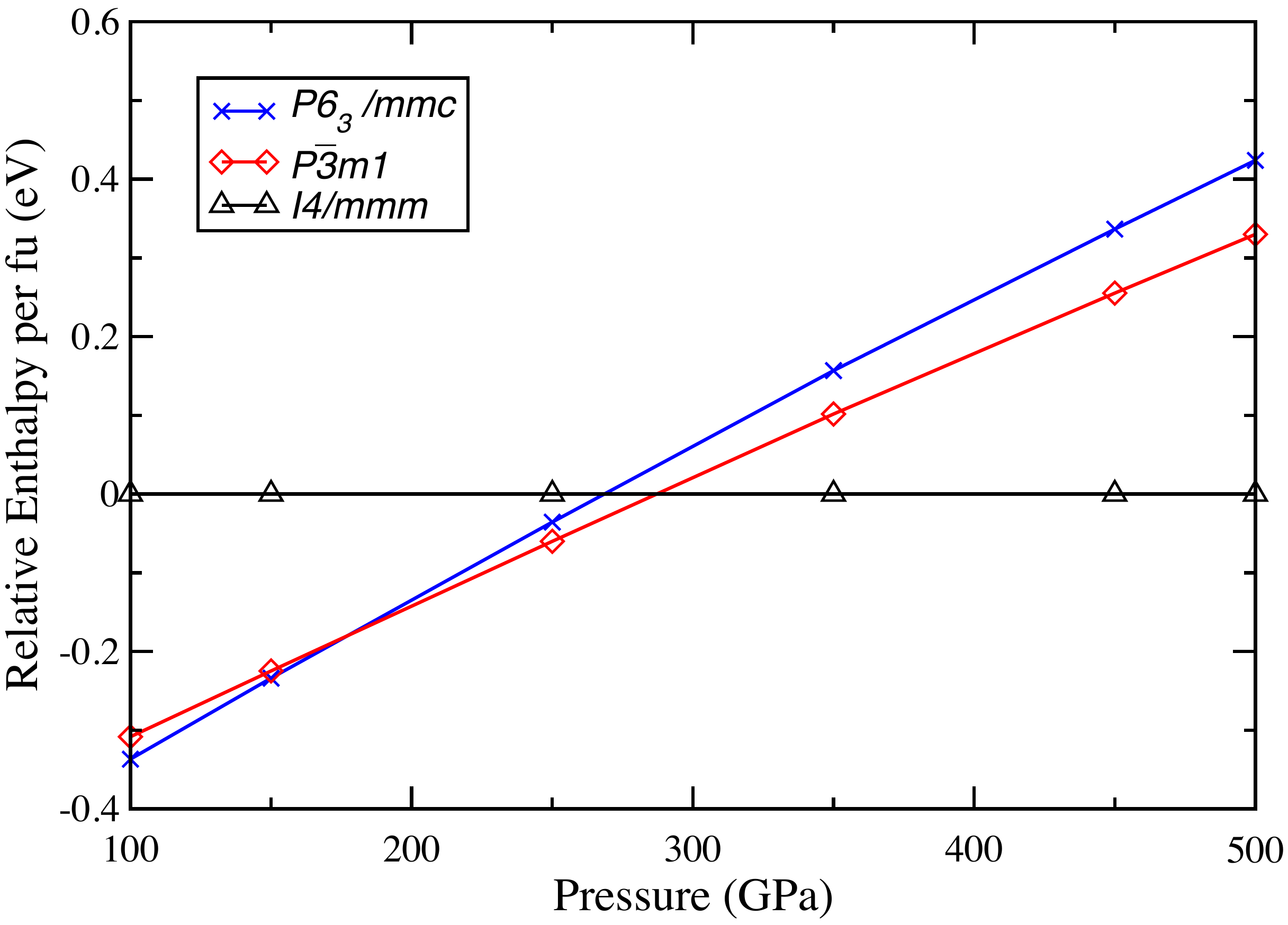} }
  \subfloat[Fe$_3$O$_2$]{
  \label{fig:ptplots:fe3o2}
  \includegraphics[width=.31\linewidth]{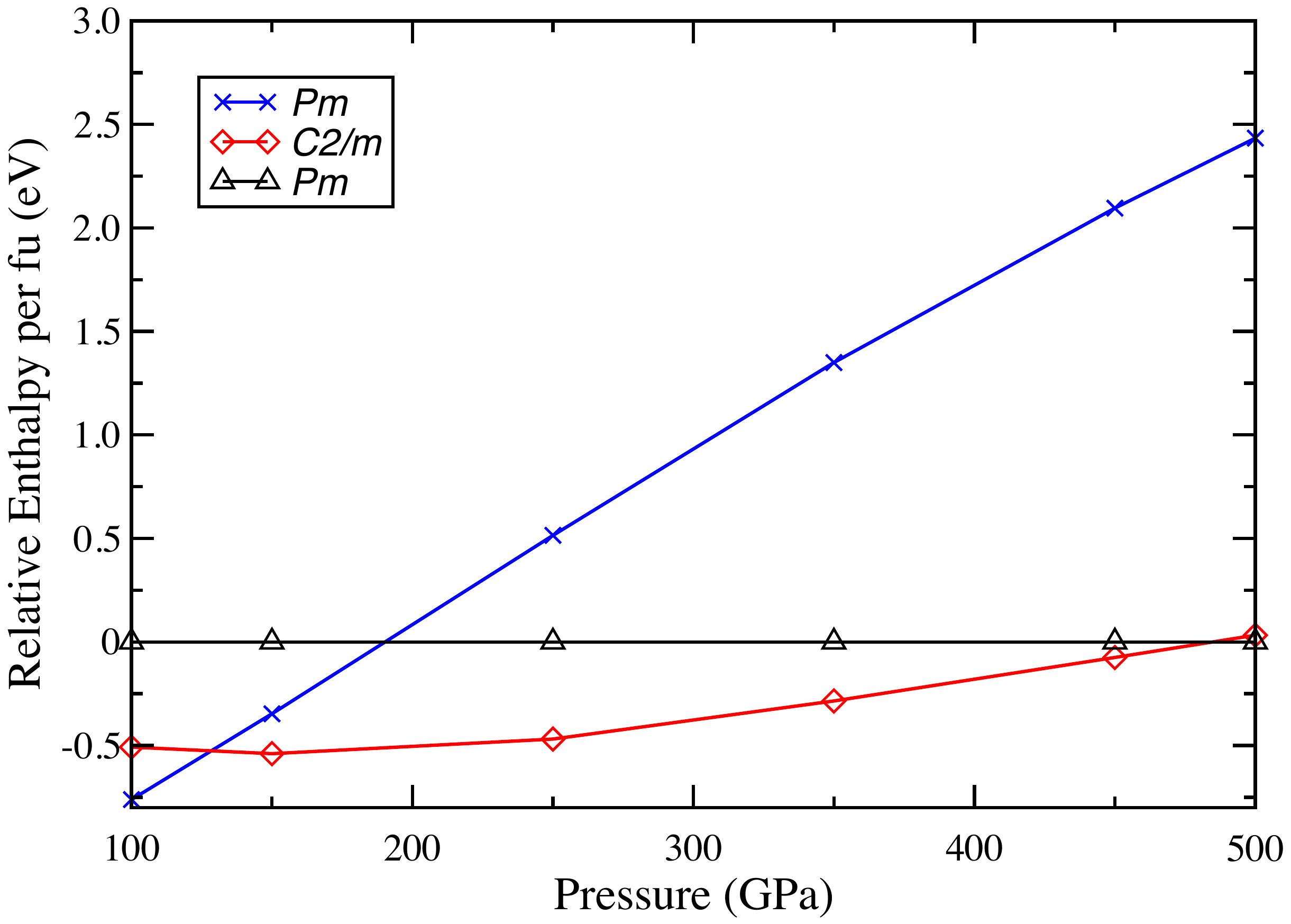} } 
  \\
  \subfloat[FeO]{
  \label{fig:ptplots:feo}
  \includegraphics[width=.31\linewidth]{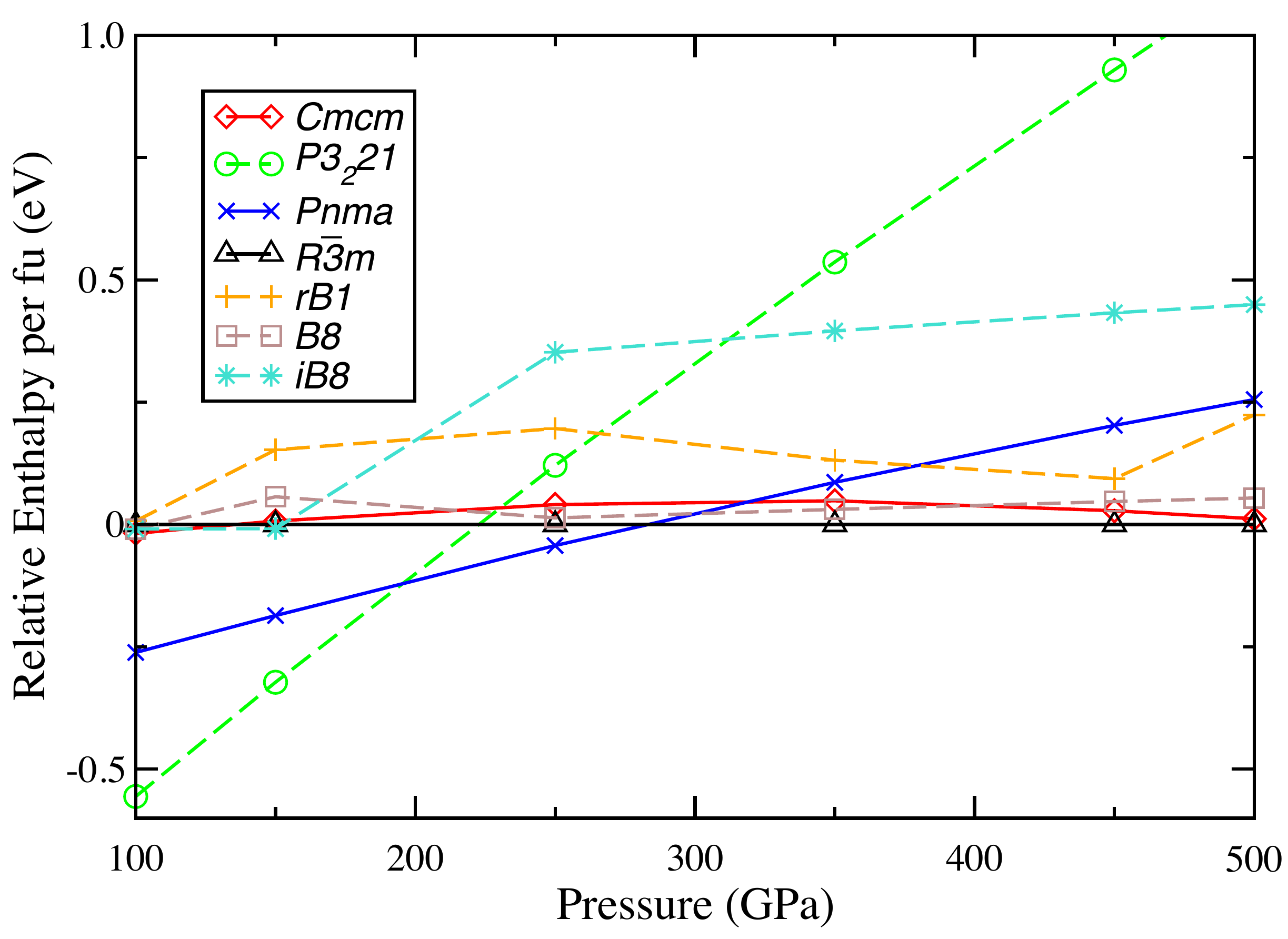} }
   \subfloat[Fe$_4$O$_5$]{
  \label{fig:ptplots:fe4o5}
  \includegraphics[width=.31\linewidth]{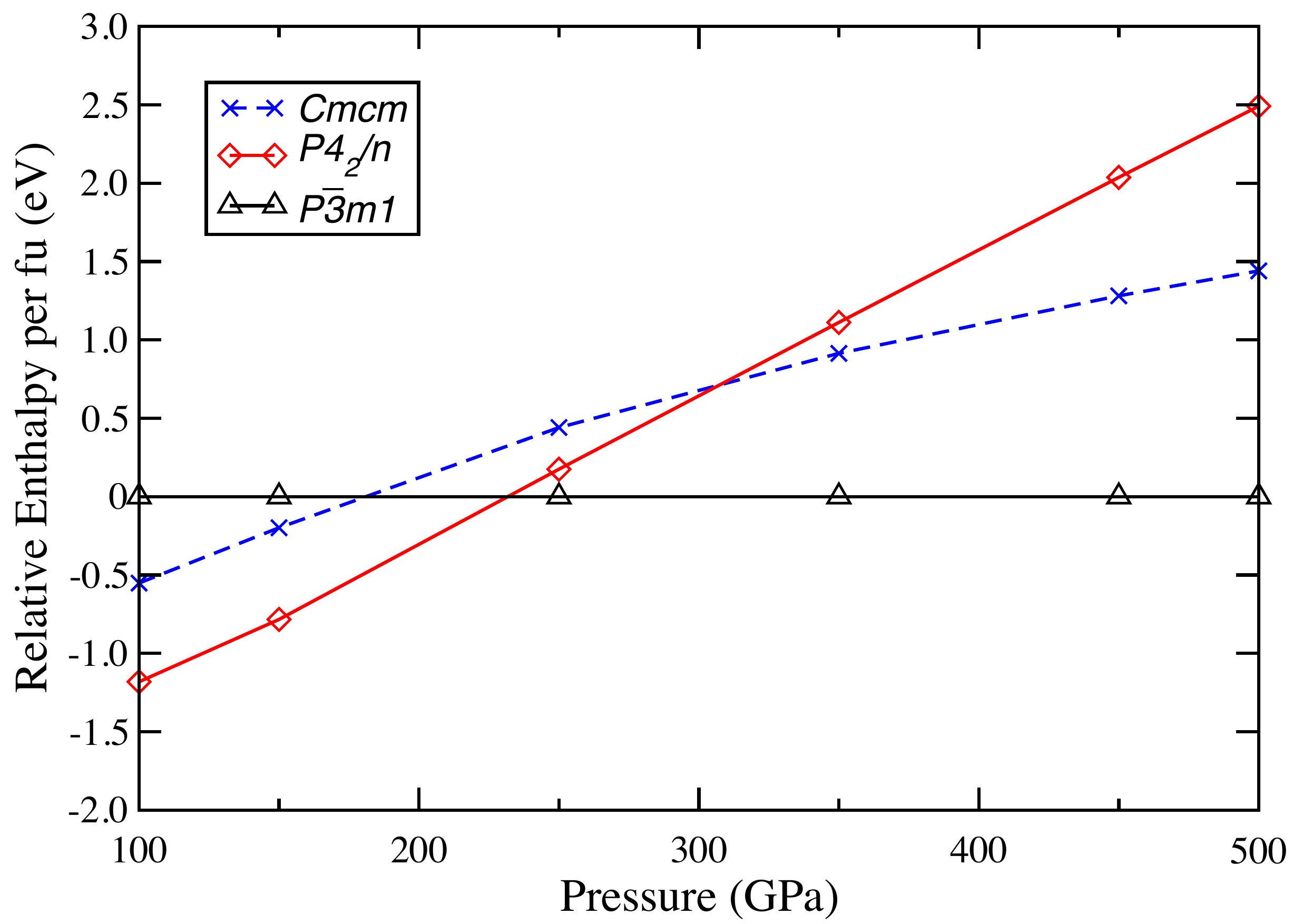} } 
  \subfloat[Fe$_3$O$_4$]{
  \label{fig:ptplots:fe3o4}
  \includegraphics[width=.31\linewidth]{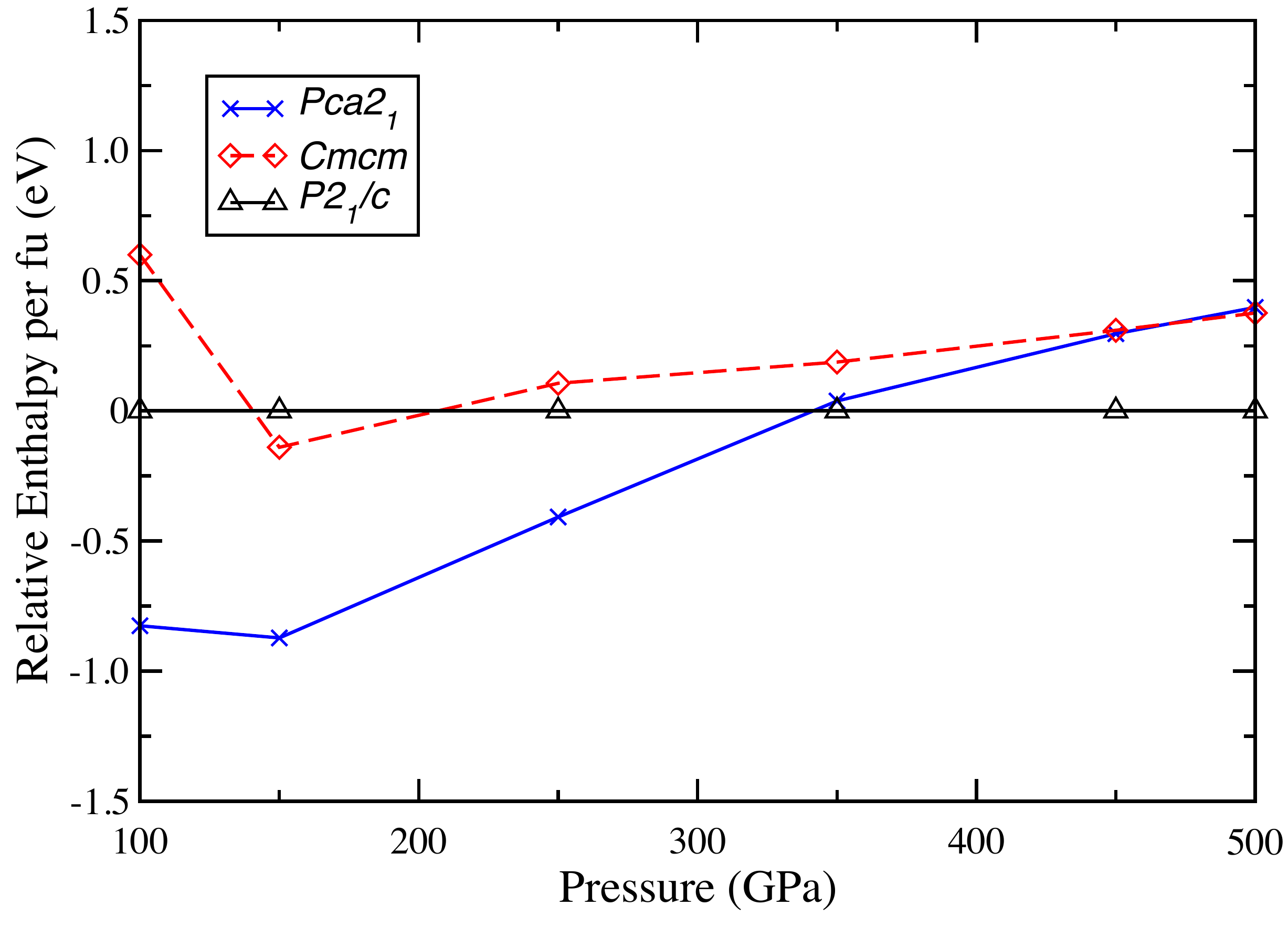} } 
  \\
   \subfloat[Fe$_2$O$_3$]{
  \label{fig:ptplots:fe2o3}
  \includegraphics[width=.31\linewidth]{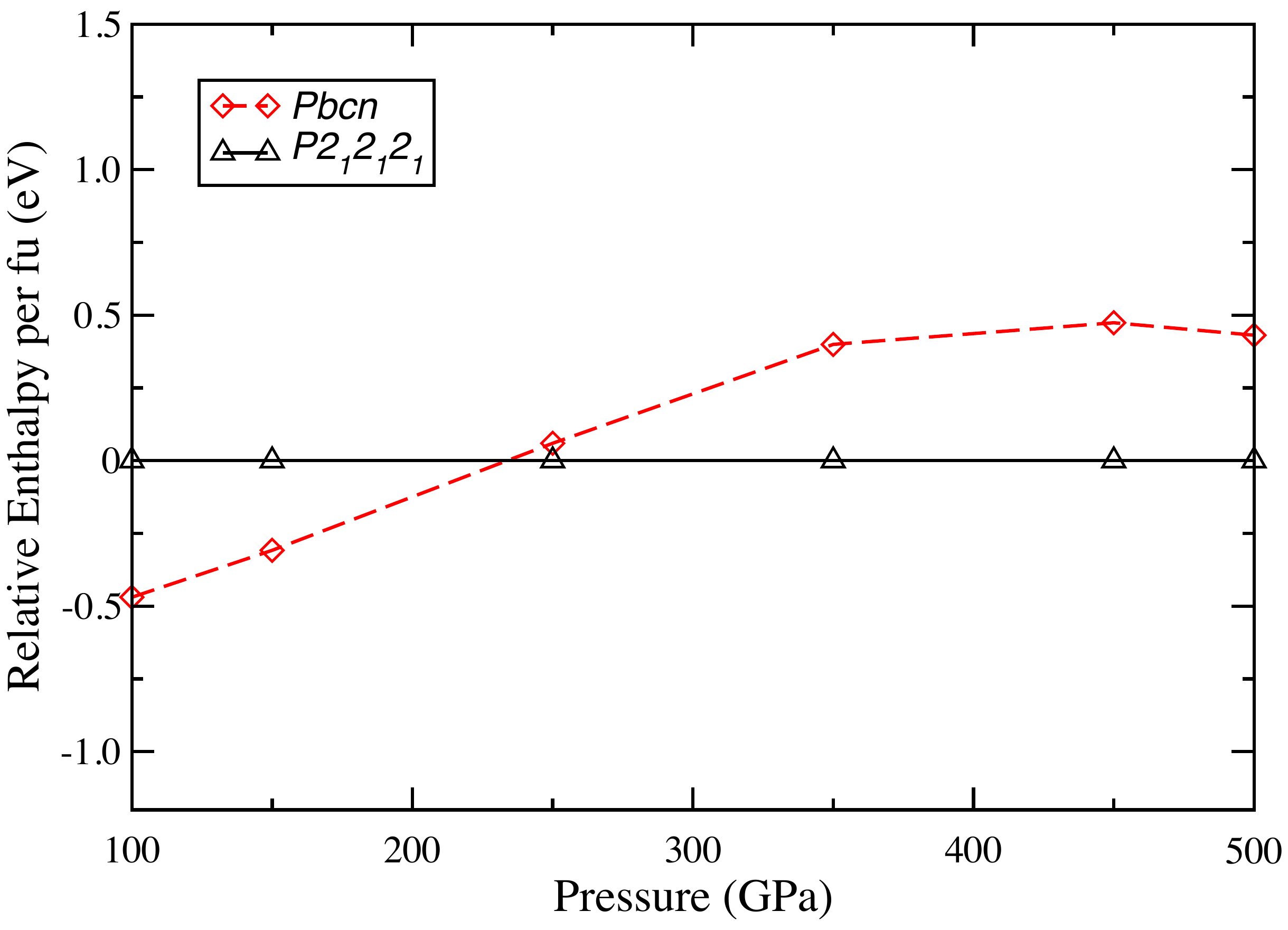} } 
   \subfloat[FeO$_2$]{
   \label{fig:ptplots:feo2}
  \includegraphics[width=.31\linewidth]{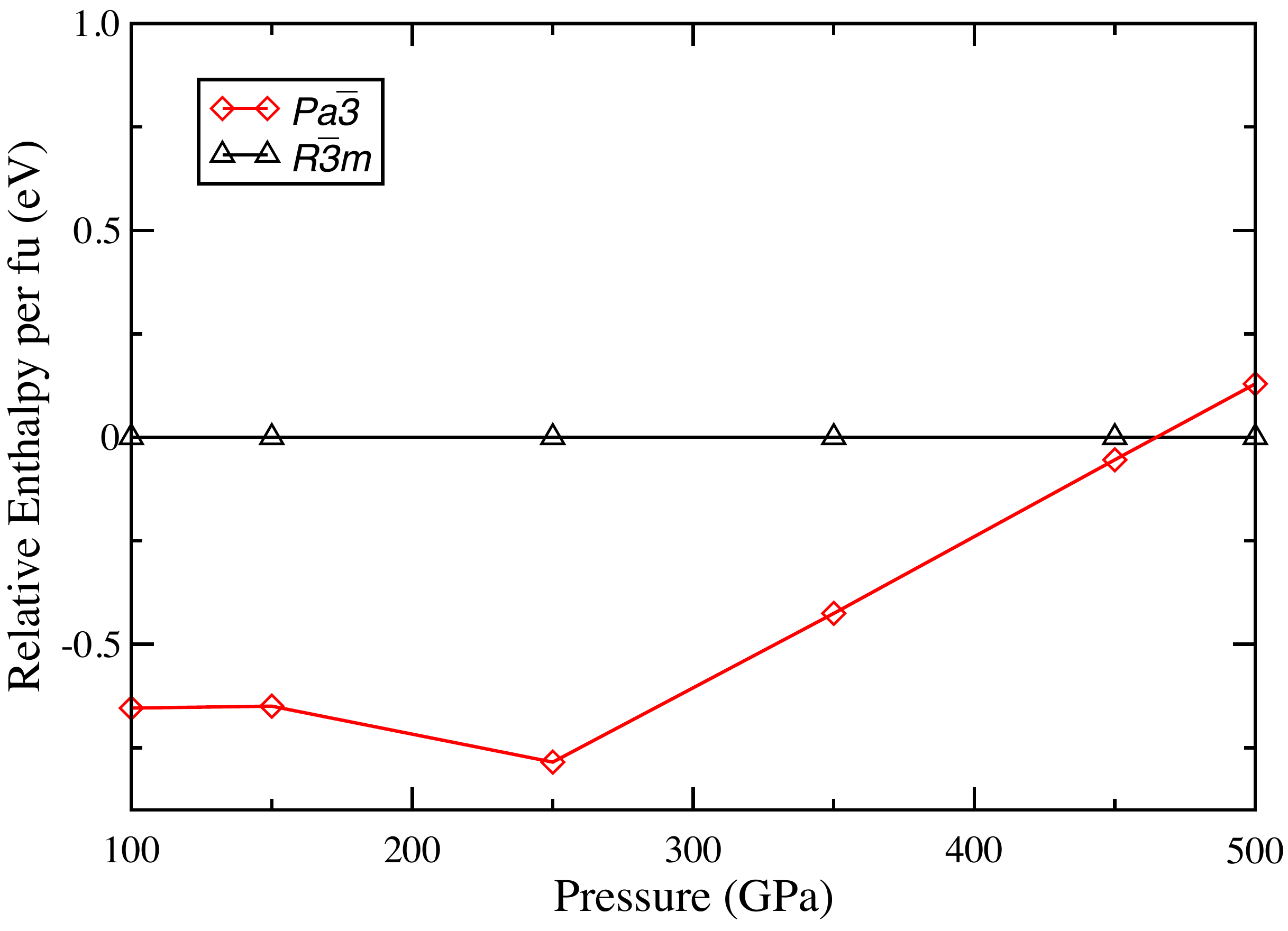} }
  \subfloat[Fe$_3$O$_7$]{
  \label{fig:ptplots:fe3o7}
  \includegraphics[width=.31\linewidth]{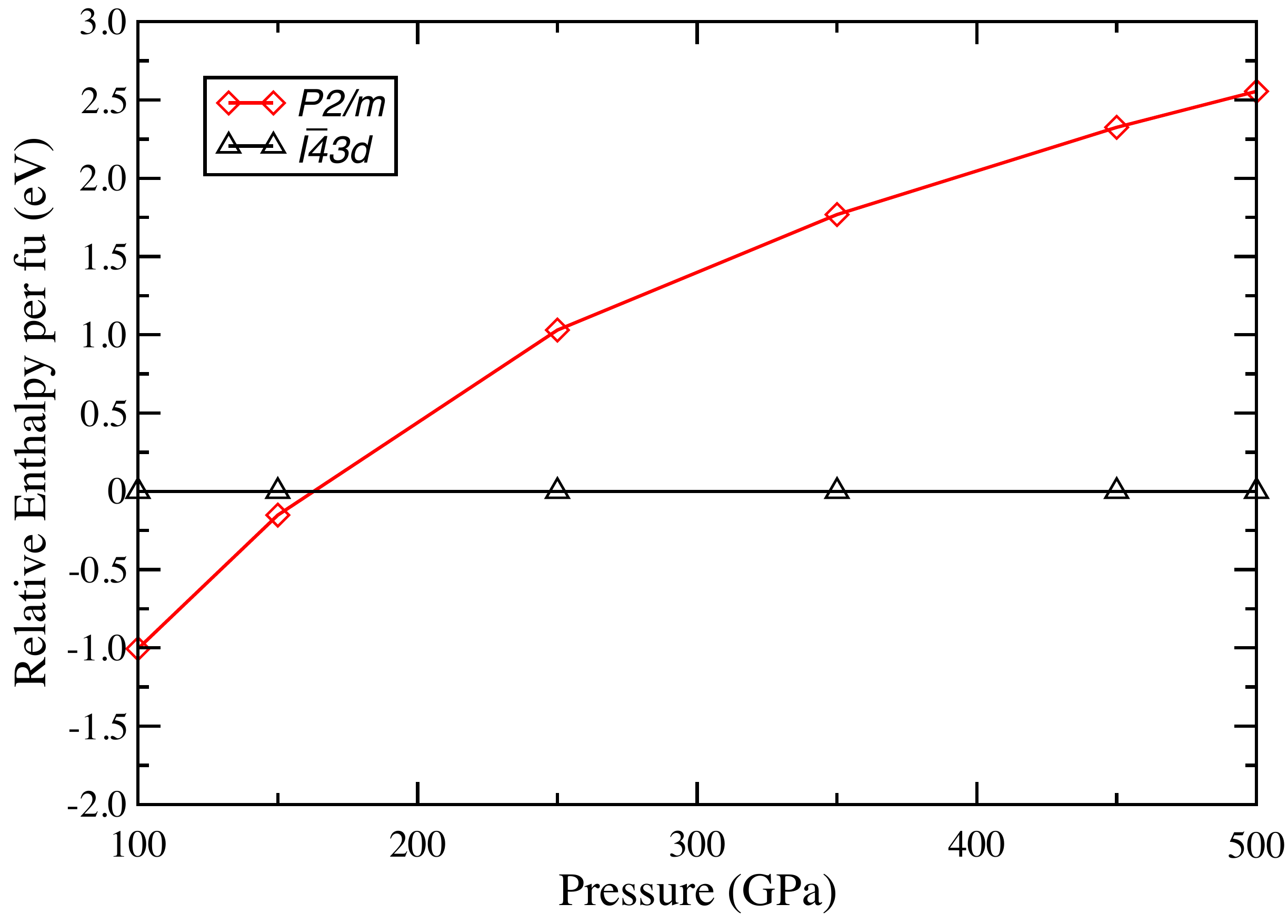} } 
  \\
  \subfloat[FeO$_3$]{
  \label{fig:ptplots:feo3}
  \includegraphics[width=.31\linewidth]{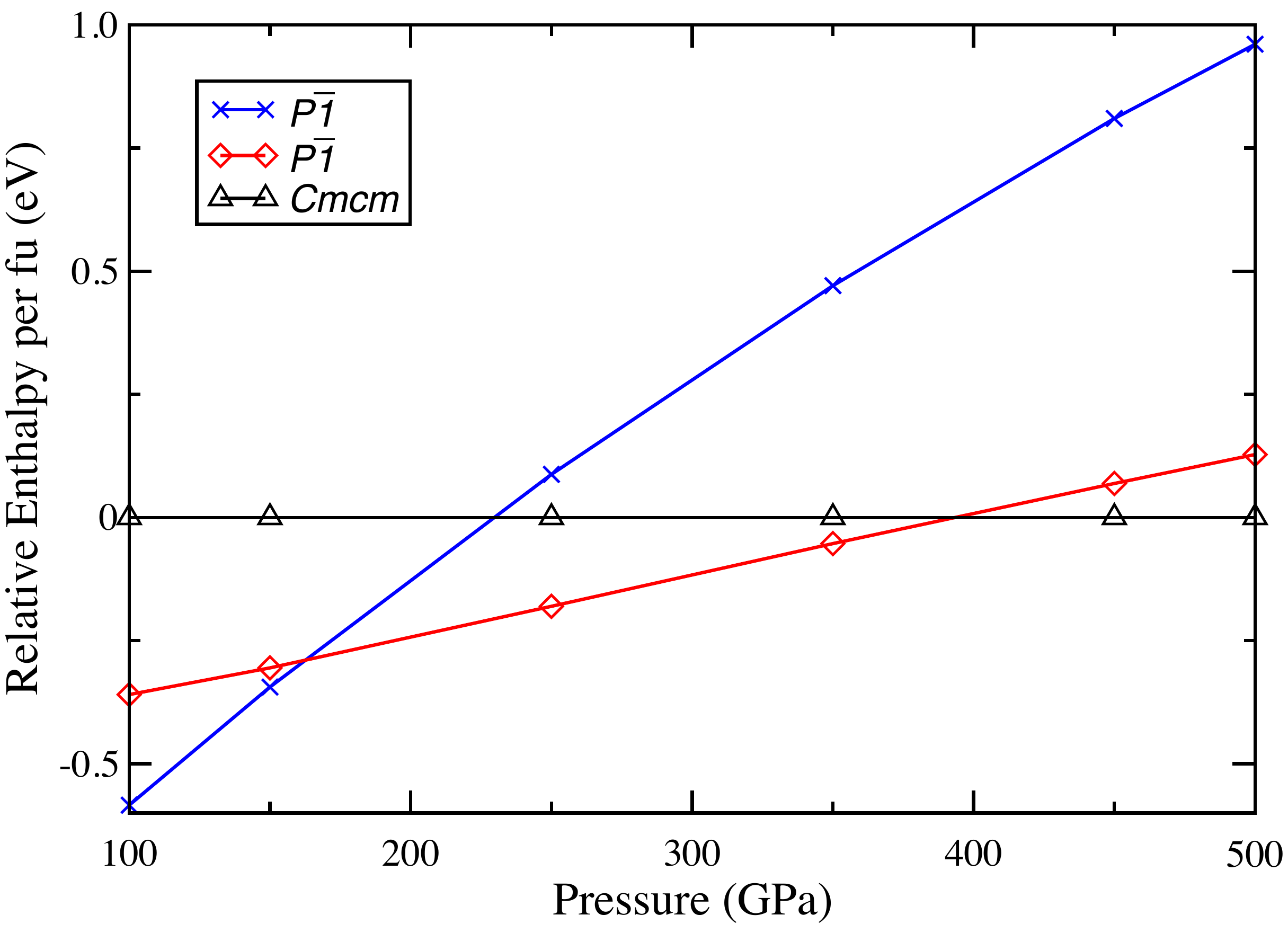} } 
  \subfloat[FeO$_4$]{
  \label{fig:ptplots:feo4}
  \includegraphics[width=.31\linewidth]{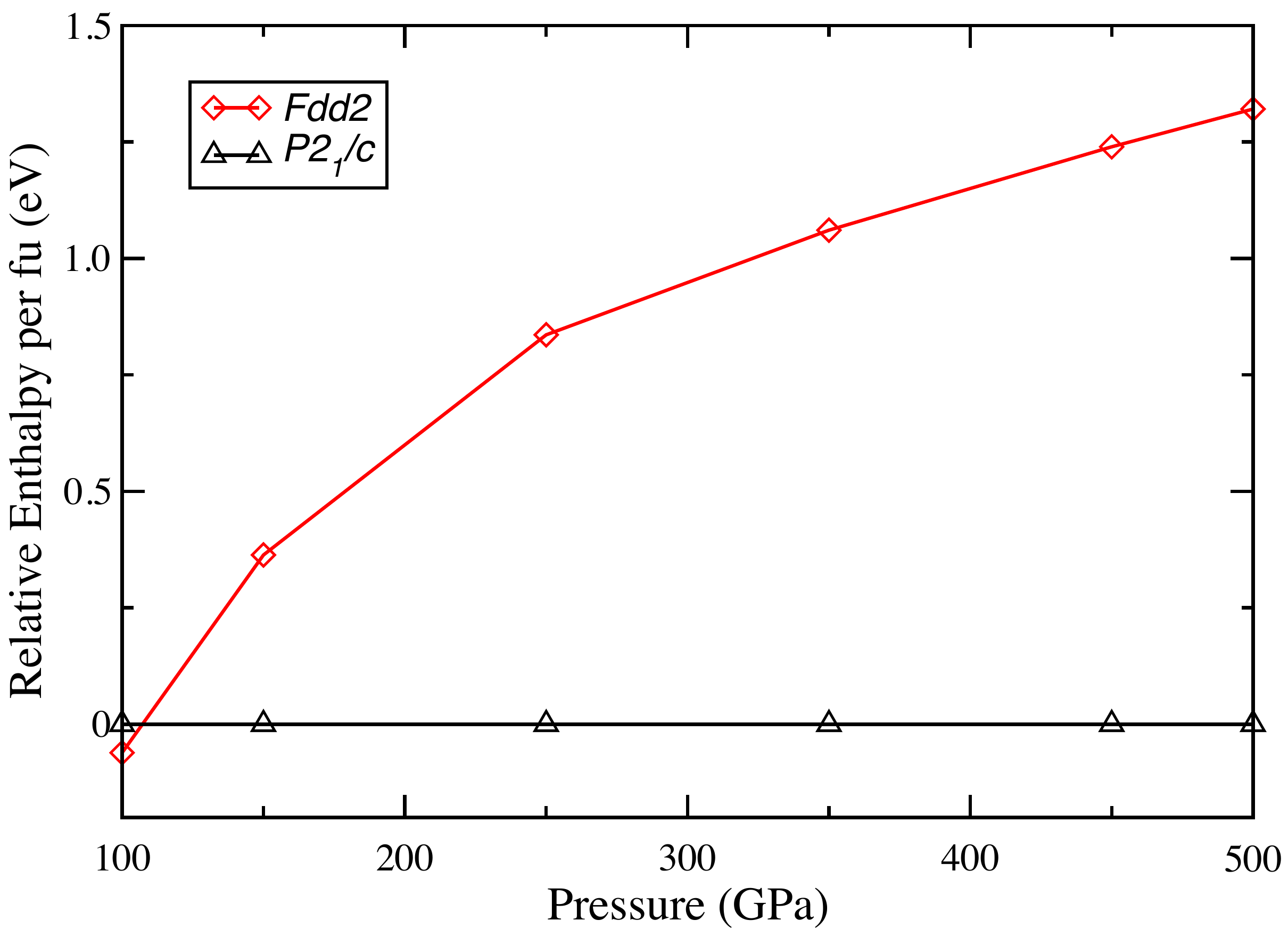} } 
\caption{\label{fig:ptplots} (Color online). Plots of relative
  enthalpy against pressure for various stoichiometries.  The
  reference structures appear as a horizontal line at 0 eV.  Dashed lines represent structures known prior to this work.}
\end{figure*} 

We predict pressure-driven phase transitions for all of the
stoichiometries considered in this study, with the exception of
Fe$_9$O.  We have obtained accurate transition pressures by performing
additional calculations at pressures around the values predicted using
Eq.\ (\ref{taylorenth}).  Plots of relative the enthalpies as a
function of pressure are shown in Fig.\ \ref{fig:ptplots}.  Three of
the stoichiometries considered, Fe$_2$O$_3$, FeO and Fe$_3$O$_4$, have
already been investigated extensively and their properties,
particularly at ambient pressure, are well known.  However, most of
the stoichiometries that we discuss here have not been studied before
and hence little is known about their high pressure behaviour.  Unless
stated otherwise, all of the lowest enthalpy phases were found by
searching and, to the best of our knowledge, have not been discovered
before.

\textit{Fe$_9$O}. The most stable structure of Fe$_9$O found has
$Amm2$ symmetry, but it turns out to be a mixture of phases.

\textit{Fe$_3$O}. See Fig.\
\ref{fig:ptplots}\subref{fig:ptplots:fe3o}.  Between 100 and 350 GPa,
the most stable structure of Fe$_3$O of $P\bar{6}m$2 symmetry turns
out to correspond to a mixture of phases rather than a single
phase. At about 350 GPa we find a phase of \textit{P4/nmm} symmetry to
be the most stable, but it also turns out to be a mixture of
phases. We also relaxed Fe$_3$O between 100 and 500 GPa in the BiI$_3$ structure, which has
previously been suggested as a low enthalpy
phase.\cite{oxygen_alfe_price} However, we found this structure to be
between 0.3 eV and 2 eV per fu higher in enthalpy than our best
structures found from searching.

\textit{Fe$_2$O}. See Fig.\
\ref{fig:ptplots}\subref{fig:ptplots:fe2o}.  We predict that Fe$_2$O
forms a structure of \textit{P}6$_3$/\textit{mmc} symmetry at 100 GPa.
At about 180 GPa we find a transformation to a phase of
\textit{P}$\bar{3}$\textit{m}1 symmetry, which is stable up to about
288 GPa, at which point we predict it to transform into a phase of
\textit{I4/mmm} symmetry.

\textit{Fe$_3$O$_2$}. See Fig.\
\ref{fig:ptplots}\subref{fig:ptplots:fe3o2}.  At 100 GPa we predict
that Fe$_3$O$_2$ will adopt a structure of $Pm$ symmetry.  At about
127 GPa, we predict a phase transition to a structure of $C2/m$
symmetry followed by a further transition to a phase of $Pm$ symmetry
at about 486 GPa.

\textit{FeO}. See Fig.\ \ref{fig:ptplots}\subref{fig:ptplots:feo}.
Previous DFT calculations for FeO have shown the stability of a phase
of $P3_221$ symmetry\cite{oganov_uspex} between 65 GPa and 190 GPa.
We also find this structure from searching and our results shown in
Fig.\ \ref{fig:ptplots}\subref{fig:ptplots:feo} agree well with those
previously reported.  However, Oganov \textit{et al.}
\cite{oganov_uspex} predict that this phase will transform to the
distorted NaCl structure at 190 GPa.  We instead find that the
$P3_221$ structure transforms to a new phase of $Pnma$ symmetry at
about 195 GPa, which then undergoes a transformation to another new
phase of $R\bar{3}m$ symmetry at about 285 GPa.  In addition to this,
Fig.\ \ref{fig:ptplots}\subref{fig:ptplots:feo} shows another new
phase of $Cmcm$ symmetry, which we predict to be stable at pressures
above 500 GPa.  Our results suggest a somewhat different picture from
that obtained from compression experiments up to 324 GPa
\cite{ozawa_feo} and at temperatures up to 4880 K. These differences
may arise for a number of reasons.  Firstly, experiments on FeO
normally use a non-stoichiometric compound, Fe$_{1-x}$O, where $x$ is
typically about 0.05, whereas our calculations assume stoichiometric
FeO.  Secondly, we have performed static lattice calculations in
contrast to real-world experiments which may yield phases stabilized
by nuclear vibrational motion.  Thirdly, kinetic barriers between
structures could also influence the structures that are observed in
experiments.  Finally, the use of an approximate density functional
will lead to further uncertainties in our results.  It is worth
remarking, however, that the PBE density functional, which obeys the
uniform limit and gives a good account of the linear response of the
electron gas to an external potential,\cite{PerdewBE96} is likely to
become more accurate as the pressure is increased.

A notable example of where our predictions for FeO differ from
observations is at relatively low pressures where experiments
(extrapolated to 0 K) have shown that FeO adopts the rhombohedrally
distorted NaCl (rB1) structure, which is predicted to be stable up to
about 100 GPa.\cite{ozawa_feo} Above 100 GPa a phase transition to the
NiAs (B8) structure is predicted. We find the inverse-B8 (iB8)
structure to be more stable than B8 between about 100 GPa and 165 GPa.
The stability of the iB8 structure has been predicted theoretically
before \cite{feo_ib8,FeO_FP_HP}, although it has yet to be verified by
experiment.  Although our results correctly predict the B8 phase to be
more stable than the rB1 phase at high pressures, our best structures
obtained from searching are lower in enthalpy than the previously
known phases across the entire pressure range considered.  We are not
aware of any experimental evidence suggesting that a $P3_221$ phase of
FeO is stable at 100 GPa, but we believe that this discrepancy may be
explained by our arguments in the previous paragraph, and we note that
earlier work suggested that the predicted stability of the $P3_221$
phase may simply be an artefact of the PBE density
functional.\cite{oganov_uspex}

\textit{Fe$_4$O$_5$}. See Fig.\
\ref{fig:ptplots}\subref{fig:ptplots:fe4o5}.  Compression experiments
have identified a phase of Fe$_4$O$_5$ with $Cmcm$ symmetry to be the
most stable between 5 and 30 GPa. \cite{lavina2011fe4o5} Our results
show this phase to be unstable between 100 and 500 GPa with respect to
the new structures found in our searches.  At 100 GPa we predict
Fe$_4$O$_5$ to adopt a structure with $P4_2/n$ symmetry prior to a
transition to a structure of $P\bar{3}m1$ symmetry at about 231 GPa,
which we predict to be stable to at least 500 GPa.

\textit{Fe$_3$O$_4$}. See Fig.\
\ref{fig:ptplots}\subref{fig:ptplots:fe3o4}.  Under ambient conditions
Fe$_3$O$_4$ adopts an inverse spinel structure of $Fd\bar{3}m$
symmetry.\cite{Ju_magnetite_2012} Between $\sim$$20-60$ GPa,
Fe$_3$O$_4$ is observed to make a slow transformation to a new phase
(sometimes denoted by h-Fe$_3$O$_4$) with orthorhombic
symmetry.\cite{Ju_magnetite_2012,dubrovinsky2003structure} The exact
structure of h-Fe$_3$O$_4$ has, however, been difficult to
resolve.\cite{dubrovinsky2003structure} In light of this, a number of
candidate structures have been proposed.  Fei \textit{et al.}\
proposed a structure of $Pbcm$ symmetry\cite{fei1999insitu} which,
although it agreed well with the available X-ray diffraction data, was
later found to be inconsistent with M\"{o}ssbauer spectroscopy
data.\cite{dubrovinsky2003structure} It was later suggested that a
structure of $Cmcm$ symmetry (sometimes referred to using the
non-standard $Bbmm$ setting) is also consistent with the X-ray
diffraction data and it currently seems to be favoured as the high
pressure phase of Fe$_3$O$_4$.\cite{haavik2000equation} First
principles calculations have shown the $Cmcm$ phase to be more stable
than the $Pbcm$ phase at pressures $>65$ GPa.\cite{Ju_magnetite_2012}
Our searches at 100 GPa found a new phase with $Pca2_1$ symmetry to be
the lowest in enthalpy up to $\sim 340$ GPa, above which a phase
transition is found to occur to another new phase of $P2_1/c$
symmetry.  Figure \ref{fig:ptplots}\subref{fig:ptplots:fe3o4} shows
that the $Cmcm$ phase is energetically unfavourable at the pressures
considered. The $Pca2_1$ structure is found to be much more stable
than the $Cmcm$ phase, being $\sim 1.4$ eV per fu lower in enthalpy at
100 GPa.  The fact that the $Pca2_1$ phase is also of orthorhombic
symmetry does not preclude it from being a candidate for
h-Fe$_3$O$_4$. Ascertaining whether the $Pca2_1$ phase is indeed
h-Fe$_3$O$_4$ would require a more detailed investigation, which we
leave to subsequent work.

\textit{Fe$_2$O$_3$}.  See Fig.\
\ref{fig:ptplots}\subref{fig:ptplots:fe2o3}. At ambient pressures the
stable phase of Fe$_2$O$_3$ is hematite, which is a corundum-type
structure of $R\bar{3}c$ symmetry.\cite{hybrid_dft_fe2o3} The high
pressure phase of Fe$_2$O$_3$ has been proposed to be either a
Rh$_2$O$_3$(II)-type structure (of \textit{Pbcn} symmetry) or a
GaFeO$_3$ orthorhombic perovskite structure.\cite{hybrid_dft_fe2o3}
Room temperature compression experiments have shown that hematite
transforms to the Rh$_2$O$_3$(II)-type structure at about 50
GPa.\cite{rt_fe2o3_compression} However, compression experiments at
higher temperatures (800-2500 K) predict a phase transition to occur
between hematite and the high pressure phase at a much lower pressure
of $\sim 26$ GPa.\cite{Ono_insitu_xray} Ono \textit{et al.}\ have
suggested that the observed transition at room temperature is between
hematite and metastable Rh$_2$O$_3$(II).\cite{Ono_insitu_xray,
  hybrid_dft_fe2o3} The perovskite-type phase is suggested to be
kinetically inhibited and therefore only obtainable at high
temperatures.\cite{Ono_insitu_xray, hybrid_dft_fe2o3} This view is
supported by hybrid-DFT calculations, which show that the
Rh$_2$O$_3$(II) phase is metastable with respect to the perovskite
phase at 50 GPa.\cite{hybrid_dft_fe2o3} At about 60 GPa and
temperatures $>1200$ K the high pressure phase is observed to
transform into a CaIrO$_3$-type (post-perovskite)
structure.\cite{hybrid_dft_fe2o3, ono2005situ} Hybrid-DFT calculations
have predicted the anti-ferromagnetically ordered CaIrO$_3$ type
structure to be the most stable between about 46 and 90 GPa with the
low-spin Rh$_2$O$_3$(II) phase gradually becoming stable as the
pressure approaches 90 GPa.  \cite{hybrid_dft_fe2o3} It appears likely
that this phase will become stable at around 100 GPa and hence become
consistent with our results shown in Fig.\
\ref{fig:ptplots}\subref{fig:ptplots:fe2o3}. We found the
(non-magnetic) Rh$_2$O$_3$(II) structure of space group \textit{Pbcn}
in our searches at 100 GPa. Our calculations find this structure to be
more stable than both the CaIrO$_3$ and GaFeO$_3$ (non-magnetic)
structures between 100 and about 233 GPa.  We predict that the
Rh$_2$O$_3$(II) phase transforms into a new phase of $P2_12_12_1$
symmetry at 233 GPa.

\textit{FeO$_2$}. See Fig.\
\ref{fig:ptplots}\subref{fig:ptplots:feo2}.  At 100 GPa FeO$_2$ we
predict a structure of $Pa\bar{3}$ symmetry, see Fig.\
\ref{fig:structures}\subref{fig:structures_feo2_pa-3}, which is stable
across a wide pressure range of about 100--465 GPa.  In this structure
the Fe atoms form a face centred cubic (fcc) arrangement.  At 465 GPa
the $Pa\bar{3}$ phase is predicted to transform into a phase of
$R\bar{3}m$ symmetry.

\textit{Fe$_3$O$_7$}. See Fig.\
\ref{fig:ptplots}\subref{fig:ptplots:fe3o7}.  Between about 100 and
160 GPa we predict a structure of $P2/m$ symmetry, and above 160 GPa
we expect it to transform into a phase of $I\bar{4}3d$ symmetry.

\textit{FeO$_3$}.  See Fig.\
\ref{fig:ptplots}\subref{fig:ptplots:feo3}.  We find a structure of
$P\bar{1}$ symmetry to be the most stable above 100 GPa, and we
predict that it will transform into a different stucture of $P\bar{1}$
symmetry above 162 GPa.  At about 395 GPa this structure is predicted
to transform into a phase of $Cmcm$ symmetry.

\textit{FeO$_4$}. See Fig.\
\ref{fig:ptplots}\subref{fig:ptplots:feo4}.  We find a structure of
$Fdd2$ symmetry to be the most stable at 100 GPa, which is predicted
to transform into a structure of $P2_1/c$ symmetry at about 108 GPa.

A summary of the structures found is given in Table
\ref{table:structures}, and the details of the structures are reported
in the Supplemental Material.\cite{EPAPS}

\section{Density of electronic states}

An analysis of the electronic densities of states of the phases was
performed with the \textsc{lindos} code \cite{LinDOSManual} which uses
a linear extrapolation
method.\cite{LinearExtrapolation1,LinearExtrapolation2} A Gaussian
smearing of width 0.3 eV was applied to the density of states in all
cases.  Figure \ref{fig:alldos} shows the density of states for all of
the structures found to lie on the convex hull.  In all cases, with
the exception of FeO$_4$ at 500 GPa, a significant density of states
is observed at the Fermi energy, indicating that these structures are
metallic.

The Fe $3s$ and Fe $3p$ states lie at, respectively, $\sim$87 and
$\sim$54 eV below the Fermi energy in all Fe-O phases.  We found the O
$2s$ states to lie at $\sim$20 eV below the Fermi energy and almost
all of the occupied electronic density of states around the Fermi
energy arises from the Fe $d$ bands, except in the O-rich structures.

\section{Conclusion} 

We have used DFT methods and the AIRSS technique to identify high
pressure structures in the Fe/O system at zero temperature across a
wide range of stoichiometries.  This is a crucial preparatory stage
for modelling the Fe/O system at finite temperatures.  By constructing
the convex hull at different pressures we have determined
energetically favourable stoichiometries, some of which have not been
considered before.  Our results broaden the range of possible stable
stoichiometries considerably from those studied previously.

We have found new structures of FeO and Fe$_2$O$_3$ which are
calculated to be stable against decomposition at 350 and 500 GPa.
Structures with stoichiometries Fe$_2$O$_3$ and FeO$_2$ are found to
be stable against decomposition at each of the three pressures
studied.  Fe$_9$O, Fe$_3$O, Fe$_2$O, Fe$_3$O$_2$, FeO, Fe$_2$O$_3$ and
FeO$_2$ are found to lie on, or very close to the hull at both 350 GPa
and 500 GPa, although some of these structures are mixtures of phases,
see Table \ref{table:structures}. At 500 GPa FeO$_4$ is also found to
lie on the hull.

We have found structures of Fe$_3$O$_4$ and Fe$_4$O$_5$ at high
pressures that are more stable than those predicted previously,
although they are unstable to decomposition at the pressures studied
here.

Increasing pressure tends to stabilize Fe/O compounds with respect to
the separated elements over the range 100--500 GPa.  Increasing
pressure stabilizes Fe rich phases, an effect that is most noticeable
between 100 and 350 GPa.  There is also a stabilization of O rich
phases with increasing pressure.

The energy reduction from compound formation in Fe/O at high pressures
is much larger than in Fe/C.  For example, at 350 GPa, the DFT results
reported in Ref.\ \onlinecite{FeCAIRSS} show that the minimum in the
convex hull occurs at Fe$_2$C with an enthalpy reduction, compared
with separated Fe and C solids, of about 0.26 eV per atom, while in
Fe/O we find an enthalpy reduction of about 1.76 eV per atom for
FeO$_2$.  Our results show that Fe/O compounds are stable (or nearly
stable) over a wider range of stoichiometries at 350 GPa than Fe/C
compounds.

\section{Acknowledgements}

This work was supported by the Engineering and Physical Sciences
Research Council U.K.\ (EPSRC-GB). Computational resources were
provided by the Cambridge High Performance Computing Service.

\end{document}